\documentclass[11pt, a4paper]{article}

\usepackage[table, svgnames, xcdraw]{xcolor}

\usepackage[utf8]{inputenc}
\usepackage[english]{babel}

\usepackage{graphicx}
\usepackage{float} 

\usepackage{amsmath, amsfonts, mathtools}

\usepackage[separate-uncertainty=true]{siunitx}

\usepackage{tabularx, booktabs, multirow}

\usepackage[hidelinks]{hyperref}
\usepackage[english, nameinlink, capitalise]{cleveref}

\usepackage{scalerel}
\usepackage{tikz}
\usetikzlibrary{svg.path}

\usepackage{geometry}
\usepackage[caption=false,font=normalsize,labelfont=sf,textfont=sf]{subfig}

\usepackage[backend=bibtex, style=ieee, sorting=none, doi=true, url=false, isbn=false, style=numeric-comp]{biblatex}
\usepackage[nolist]{acronym}

\usepackage{textcomp}
\usepackage{enumitem}
\usepackage{balance}

\usepackage{fancyhdr}
\definecolor{orcidlogocol}{HTML}{A6CE39}
\tikzset{
  orcidlogo/.pic={
    \fill[orcidlogocol] svg{M256,128c0,70.7-57.3,128-128,128C57.3,256,0,198.7,0,128C0,57.3,57.3,0,128,0C198.7,0,256,57.3,256,128z};
    \fill[white] svg{M86.3,186.2H70.9V79.1h15.4v48.4V186.2z}
                 svg{M108.9,79.1h41.6c39.6,0,57,28.3,57,53.6c0,27.5-21.5,53.6-56.8,53.6h-41.8V79.1z M124.3,172.4h24.5c34.9,0,42.9-26.5,42.9-39.7c0-21.5-13.7-39.7-43.7-39.7h-23.7V172.4z}
                 svg{M88.7,56.8c0,5.5-4.5,10.1-10.1,10.1c-5.6,0-10.1-4.6-10.1-10.1c0-5.6,4.5-10.1,10.1-10.1C84.2,46.7,88.7,51.3,88.7,56.8z};
  }
}
\newcommand{\orcidicon}[1]{\href{https://orcid.org/#1}{\mbox{\scalerel*{
\begin{tikzpicture}[yscale=-1,transform shape]
\pic{orcidlogo};
\end{tikzpicture}
}{|}}}}

\uspunctuation

\begin{acronym}
\acro{AF}{additional features}
\acro{TDC}{time-to-digital converter}
\acro{TOF}{time-of-flight}
\acro{SNR}{signal-to-noise ratio}
\acro{COG}{center of gravity}
\acro{CTR}{coincidence time resolution}
\acro{cal}{detector under calibration}
\acro{CNN}{convolutional neural network}
\acro{coinc}{coincidence detector}
\acro{DOI}{depth of interaction}
\acro{GTB}{gradient tree boosting}
\acro{GBDT}{gradient-boosted decision trees}
\acro{IR}{isotonic regression}
\acro{kNN}{$k$ nearest neighbors}
\acro{LSF}{line spread function}
\acro{ML}{Machine Learning}
\acro{MAE}{mean absolute error}
\acro{LYSO}{lutetium–yttrium oxyorthosilicate}
\acro{LSO}{Lutetium oxyorthosilicate}
\acro{SR}{spatial resolution}
\acro{SD}{slab detector}
\acro{OD}{one-to-one detector}
\acro{SPAD}{single photon avalanche diode}
\acro{PCA}{principal component analysis}
\acro{PDPC}{Philips Digital Photon Counting}
\acro{PSF}{point spread function}
\acro{RR}{retroreflector}
\acro{PET}{positron emission tomography}
\acro{FWHM}{full width at half maximum}
\acro{ESR}{enhanced specular reflector}
\acro{MLITE}{maximum likelihood interaction time estimation}
\acro{TDL}{tapped delay line}
\acro{LOR}{line-of-response}
\acro{BGO}{Bismuth germanate}
\acro{SiPM}{silicon photomultiplier}
\acro{PNGD}{percentage of non-gaussian prediction distributions}
\acro{AI}{artificial intelligence}
\acro{GBDT}{gradient-boosted decision tree}
\acro{FPGA}{field programmable gate array}
\acro{MR}{memory requirement}
\end{acronym}

\newcommand{\scBroad}{0.48}
\newcommand{\scBroadd}{0.6}
\newcommand{\mygray}{gray!50} 

\begin{document}
\title{\textbf{Rethinking Timing Residuals: Advancing PET Detectors with Explicit TOF Corrections}}
\author{Stephan Naunheim$^{1,2,*}$\orcidicon{0000-0003-0306-7641}, Luis Lopes de Paiva$^{1,2}$\orcidicon{0009-0004-4944-2720}, Vanessa Nadig$^{2}$\orcidicon{0000-0002-1566-0568}, Yannick Kuhl$^{1,2}$\orcidicon{0000-0002-4548-0111},\\
Stefan Gundacker$^{2,3}$\orcidicon{0000-0003-2087-3266}, Florian Mueller$^{2}$\orcidicon{0000-0002-9496-4710}, Volkmar Schulz$^{1,4,5,6,*}$\orcidicon{0000-0003-1341-9356}
\thanks{
$^{1}$Institute of Imaging and Computer Vision (LfB), RWTH Aachen University, Aachen, Germany\\
$^{2}$Department of Physics of Molecular Imaging Systems (PMI), Institute for Experimental Molecular Imaging, RWTH Aachen University, Aachen, Germany\\
$^{3}$Institute of High Energy Physics (HEPHY), Austrian Academy of Sciences, Vienna, Austria\\
$^{4}$Hyperion Hybrid Imaging Systems GmbH, Aachen, Germany \\
$^{5}$Fraunhofer Institute for Digital Medicine MEVIS, Aachen, Germany \\
$^{6}$Physics Institute III B, RWTH Aachen University, Aachen, Germany\\
$^{*}$Corresponding Authors: Stephan Naunheim, Volkmar Schulz\\
$\{$stephan.naunheim, volkmar.schulz$\}$@lfb.rwth-aachen.de}}
\date{}

\maketitle
\thispagestyle{fancy}

\begin{abstract}
PET is a functional imaging method that can visualize metabolic processes and relies on the coincidence detection of emitted annihilation quanta. From the signals recorded by coincident detectors, TOF information can be derived, usually represented as the difference in detection timestamps. Incorporating the TOF information into the reconstruction can enhance the image’s SNR. Typically, PET detectors are assessed based on the coincidence time resolution (CTR) they can achieve. However, the detection process is affected by factors that degrade the timing performance of PET detectors. Research on timing calibrations develops and evaluates concepts aimed at mitigating these degradations to restore the unaffected timing information. While many calibration methods rely on analytical approaches, machine learning techniques have recently gained interest due to their flexibility. We developed a residual physics-based calibration approach, which combines prior domain knowledge with the flexibility and power of machine learning models. This concept revolves around an initial analytical calibration step addressing first-order skews. In the subsequent step, any deviation from a defined expectation is regarded as a residual effect, which we leverage to train machine learning models to eliminate higher-order skews. The main advantage of this idea is that the experimenter can guide the learning process through the definition of the timing residuals. In earlier studies, we developed models that directly predicted the expected time difference, which offered corrections only implicitly (implicit correction models). In this study, we introduce a new definition for timing residuals, enabling us to train models that directly predict correction values (explicit correction models). We demonstrate that the explicit correction approach allows for a massive simplification of the data acquisition procedure, offers exceptionally high linearity, and provides corrections able to improve the timing performance from \qty{371 \pm 6}{\pico \second} to \qty{281 \pm 5}{\pico \second} for coincidences from \qtyrange{430}{590}{\kilo \eV}. Furthermore, the novel definition makes it possible to exponentially reduce the models in size, making it suitable for applications with high data throughput, such as PET scanners. All experiments are performed with two detector stacks comprised of \numproduct{4x4} LYSO:Ce,Ca crystals (each \qtyproduct{3.8 x 3.8 x 20}{\milli \metre}), which are coupled to \numproduct{4x4} Broadcom NUV-MT SiPMs and digitized with the TOFPET2 ASIC.\newline

\end{abstract}

\acresetall
\section{Introduction}
\label{sec:intro}

The introduction of precise \ac{TOF} information in \ac{PET} leads to a significant improvement in the \ac{SNR} of the reconstructed images, which could aid physicians in diagnosis \cite{karp_benefit_2008,kadrmas_impact_2009}. For this reason, in recent years, there has been increased research into various approaches that have the potential to further improve \ac{TOF}-\ac{PET}, with the ultimate goal of eventually achieving a timing resolution in the order of \qty{10}{\pico \second} \cite{lecoq_roadmap_2020}.\newline
The \ac{PET} data acquisition begins with the detection of the emitted $\gamma$-photons using dedicated detectors consisting of a scintillation volume and attached readout technology. Recent research demonstrated the potential of co-doping approaches \cite{nemallapudi_sub-100_2015,nadig_timing_2023}, Cherenkov emitters \cite{brunner_studies_2014,terragni_time_2022} like \ac{BGO} \cite{brunner_bgo_2017, kratochwil_exploring_2021, gonzalez-montoro_cherenkov_2022, cates_scintillation_2024, pourashraf_magnetic_2025} or composite materials like metascintillators \cite{konstantinou_metascintillators_2022, konstantinou_proof--concept_2023}. Furthermore, the research community investigates double-sided readout techniques \cite{seifert_improving_2015, borghi_32_2016, tabacchini_improved_2017, kratochwil_timing_2024} and waveform sampling approaches \cite{gundacker_high-frequency_2019, cates_low_2022,nadig_16-channel_2023, weindel_time-based_2023}, both providing rich signal information but posing hardware-related challenges regarding the usage in full scanner configurations.\newline
Besides previously mentioned research lines, a considerable number of studies are conducted on the development and improvement of calibration techniques aiming to minimize any deteriorating effects impacting the \ac{TOF} measurement by estimating unaffected timestamps or providing corrections to them. Mathematically, one can describe a timestamp $t_d$ reported by the detection system as a superposition of unaffected, true timestamp $\Theta$ and a time offset $\Delta t_s$ often called skew, 
\begin{equation}
t_d = \Theta + \Delta t_s,
\end{equation}
with the offset value being a composition of different deteriorating sub-offset $s_i$, which can be parameterized by a set of parameters $\vec{p_i}$,
\begin{equation}
\Delta t_s = \sum \limits_{i} s_i(\vec{p_i}).
\end{equation}
Here, $\vec{p_i}$ represents a set of parameters that governs the behavior of the sub-skew $s_i$ without loss of generality.
In reality, the sub-offset $\{ s_i\}$ might change from $\gamma$-interaction to $\gamma$-interaction (e.g., due to different energy depositions affecting the signal steepness, baseline shifts, \ac{DOI}-related effects, etc. \cite{piemonte_overview_2019, acerbi_understanding_2019}) leading to different magnitudes and constitutions of $\Delta t_s$.\newline
While in former times, primarily analytical calibration approaches were studied, \ac{AI} methods gained interest in recent years, so that machine learning approaches are also an active area of research. Classical \ac{PET} timing calibration methods in the literature rely on solving matrix equations \cite{mann_computing_2009, reynolds_convex_2011, freese_robust_2017} or use timing alignment probes \cite{thompson_central_2005, bergeron_handy_2009}. Also, data-driven approaches without requiring specialized data acquisition setups have been investigated, e.g., using consistency conditions \cite{werner_tof_2013,defrise_time--flight_2018,li_consistency_2019,li_autonomous_2021} or detector-intrinsic characteristics \cite{rothfuss_time_2013, panin_lso_2022}. Other approaches use statistical modeling under the premise of a maximum-likelihood approach \cite{dam_sub-200_2013, borghi_towards_2016}.\newline
Many machine learning-based timing calibrations utilize supervised learning and, therefore, generate labeled data by measuring radiation sources at multiple positions \cite{berg_using_2018, labella_toward_2021, feng_transformer-cnn_2024, muhashi_enhancing_2024, feng_predicting_2024} or at a single position \cite{onishi_unbiased_2022}. Nearly all of the existing approaches work with the signal waveforms. Although this data represents a rich source of information, it often requires dedicated measurement hardware capable of high sampling rates, e.g., oscilloscopes, that might be challenging to implement in a full scanner setting.\newline
In recent studies, we developed and demonstrated the functionality of a machine learning-based timing calibration for clinically relevant detectors (coupled to digital \acp{SiPM} \cite{naunheim_improving_2023, naunheim_first_2023} and analog \acp{SiPM} \cite{naunheim_holistic_2024}) without the need of waveform information. Instead, we utilized detector blocks coupled to matrices of \acp{SiPM}, as is done in pre-clinical and clinical settings. Furthermore, we proposed to follow a residual physics-based calibration concept, which separates the calibration effort into two parts.\newline
In the first part, an analytical calibration procedure is conducted, which addresses so-called skews of first order that do not change during measurement and are similar for each event. After correction for those skews, each deviation from an expectation can be interpreted as a residual effect caused by higher-order skews that might change on an event basis. To address these higher-order skews in the second part, we endeavor to use machine learning techniques since they offer high flexibility and are suitable for finding characteristic patterns in the calibration data. The strength of the residual physics concept is the underlying idea that the experimenter can incorporate prior domain knowledge in the way the residual effects are defined.\newline
When applying machine learning techniques in physical problems, one should ensure that the trained models produce results that are in alignment with physical laws. Therefore, we proposed in prior works, to use a three-folded evaluation scheme. In the first instance, models are assessed regarding their \ac{MAE} performance. How well the models are with physics is tested in the second instance. Finally, models that passed the first two evaluation levels are tested in the third instance regarding the improvement in \ac{CTR}.\newline
In our proof-of-concept study \cite{naunheim_improving_2023}, we used a similar label generation like Berg and Cherry \cite{berg_using_2018}, which demands the collection of multiple source positions between two detectors in order to label the data for supervised training of the model. Successfully trained regression models are able to predict the expected time difference for a given input sample, e.g., consisting of information about the measured time difference, the detected timestamps, the detected energy signals, and the estimated $\gamma$-interaction position. In this sense, the trained models provide implicit corrections by directly predicting the corrected time difference.\newline
Although this procedure demonstrated that significant improvements in \ac{CTR} are possible, the acquisition of the labeled data was time-consuming considering future applications in a fully equipped \ac{PET} scanner. In a follow-up work \cite{naunheim_holistic_2024}, we demonstrated that the acquisition time can be significantly reduced without losing the calibration quality making the approach more practical. However, the study also revealed a strong dependence of the trained regression models on the used training step width given as the distance between subsequent source positions. As soon as this stepping distance becomes too large, the regression model showed on finer sampled test data a form of collapse towards a classification model for positions present in the training data set. This resulted in a worse prediction quality for radiation sources located at positions being not present in the training data. Even though the overall prediction quality of the collapsing models was in an acceptable range regarding the \ac{MAE}, it also led to a substantial loss of linearity. The linearity property of a calibration model, namely that a source shift in the spatial domain translates linearly to shift of the time difference distribution in the time domain, is especially for \ac{TOF}-\ac{PET} one of the most important attributes, since it ensures a correct interpretation of the \ac{TOF} information. The precise value of the maximum training stepping width at which the model remains functional might depend on the spatial sampling of the training dataset and on the timing resolution that can be achieved with the detectors being calibrated. While the precise derivation of the dependence on the measurement parameters is out of the scope of this work, one can safely assume that a training step width being large compared to either the testing step width or/and the timing resolution will likely result in a model collapse. The implication of this observation leads to the conclusion that the smaller the \ac{CTR} value and/or testing step width is, the smaller the training step width used has to be, resulting potentially in an elaborate data acquisition.\newline
In this work, we provide a new definition of the timing residuals (oriented on the work of Onishi et al. \cite{onishi_unbiased_2022}), which allows us to train \ac{TOF}-correction models explicitly, providing correction values instead of expected time differences.\newline
In the first part, we compare the novel residual formulation (explicit corrections) with the already established one (implicit corrections) regarding a three-folded evaluation scheme. The analyses use real measurement data and demonstrate that by using the explicit correction approach, the strong dependence on the step width is removed, the biased prediction distributions are suppressed, and the linearity property is preserved. Considering a typical \ac{PET} scanner geometry, we will refer to this study as the \textit{transaxial performance} study.\newline
In the second part, we analyze how the in-plane distribution of sources for a given location between the detectors affects the calibration quality, minding a later in-system application with, e.g., dedicated phantoms or a point source located at a single position between the detectors. In the following, this study is called the \textit{in-plane distribution} study.

\section{Materials}

\subsection{PET Detectors}

Two clinically relevant \ac{PET} detectors of identical design are used for the experiments. The scintillator topology is given by \numproduct{4 x 4} LYSO:Ce,Ca crystals manufactured by Taiwan Applied Crystals (TAC), which have shown promising performance in previous studies \cite{nadig_timing_2023}. ESR-OCA-ESR sheets with a thickness of \qty{155}{\micro \metre} cover the lateral sides of the outer crystals. Each of the \numproduct{4x4} crystal elements encloses a volume of \qtyproduct{3.8 x 3.8 x 20}{\milli \metre}, features polished top and bottom faces with depolished lateral faces \cite{trummer_depth_2009,pizzichemi_light_2019}, and allows for light-sharing. The crystals are coupled to a \numproduct{4x4} array of Broadcom NUV-MT SiPMs (AFBR-S4N44P164M) having a pitch of \qty{3.8}{\milli \metre} and \qty{40}{\micro \metre} SPAD size. Each SiPM is coupled to one channel of the TOFPET2 ASIC (version 2c) \cite{bugalho_experimental_2019,nadig_asics_2025}. A timestamp in picoseconds and an energy value in arbitrary units is reported if an SiPM is triggered and defines a so-called hit. The trigger and validation settings can be found in \cref{tab:measurement_settings}. The overvoltage is set to \qty{7}{\volt} to balance good timing performance \cite{nadig_timing_2023} with a reasonable noise floor \cite{nadig_timing_2023, nadig_16-channel_2023}.

\subsection{Experimental Setup \& Settings}

The measurement setup is located in a light-protected climate chamber, which is controlled at a temperature of \qty{16}{\degreeCelsius}. The sensor tiles are mounted at a distance of \qty{565}{\milli \metre}. In between the detectors, a $^{22}$Na point source with an active diameter of \qty{0.5}{\milli \metre} is used. The source holder is connected to a custom-manufactured translation stage system, allowing movement along all three spatial axes with a precision of $< \qty{1}{\micro\metre}$. The activity of the source was given to \qty{1.4}{\mega \becquerel}. The ASIC board temperatures are kept constant while operation and are measured to be \qty{33.9 \pm 0.1}{\degreeCelsius} and \qty{36.3 \pm 0.1}{\degreeCelsius}, respectively. For all data acquisitions, the same measurement settings (listed in \cref{tab:measurement_settings}) are used.

\subsection{Defining Timing Residuals: Implicit \& Explicit Corrections}

\begin{figure}[htb]
\centering
\includegraphics[width=0.9 \textwidth]{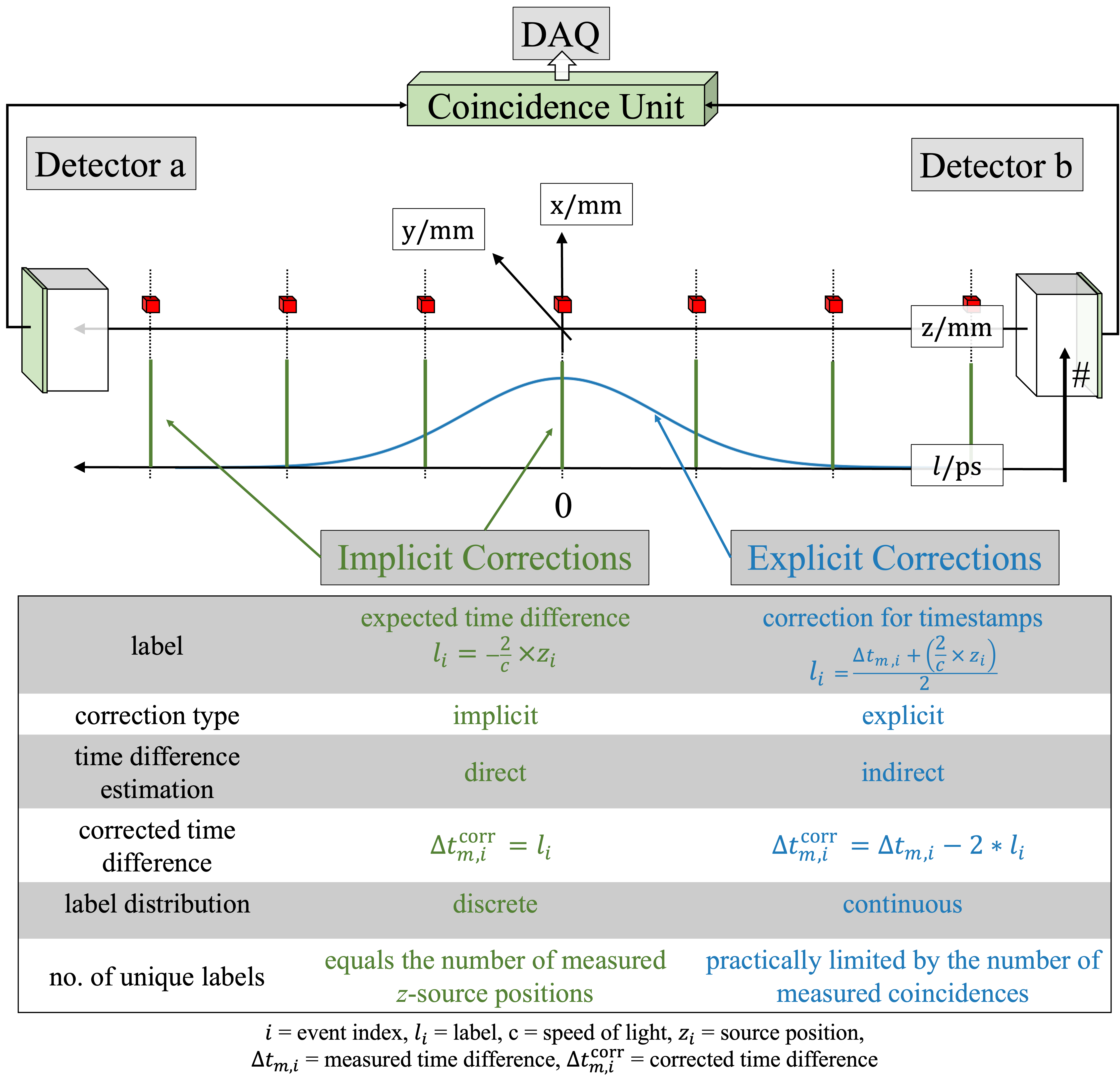}
\caption{Visualization of the label distributions of the implicit (green) and explicit (blue) correction approach. The radiation source positions are displayed as red cubes.}
\label{fig:LabelScheme}
\end{figure}

Labeling the acquired measurement data is essential to make it applicable for supervised learning algorithms. The labels represent a type of ground truth since they guide the training procedure of machine learning models. However, for \ac{PET} timing calibrations, the underlying ground truth is not accessible since the time skews can vary on an event basis, and the only information available to the experimenter is the measured time difference given as the difference between the timestamps reported by two coincident detectors. Nevertheless, it is possible to obtain a certain level of ground truth by connecting the spatial location of a radiation source with the expected time difference, assuming that no systematic skews are present. By shifting the radiator to a different spatial location $z$, one can anticipate that also the expected time difference $\Delta t_{\mathbb{E}}$ will shift. Both quantities are related to each other via, 

\begin{equation}
\Delta t_{\mathbb{E}}(z) = -\frac{2}{c} \cdot z,
\label{eq:implicitLabel}
\end{equation}
with $c$ denoting the speed of light. In our proof-of-concept work \cite{naunheim_improving_2023}, we used \cref{eq:implicitLabel} to generate labels $l_i$ by shifting a radiation source between the detectors,
\begin{equation}
l_i \coloneq \Delta t_{\mathbb{E},i}(z_i).
\end{equation}
Models trained with this definition of the timing residuals correct the measured timestamps $t_{a/b}$ of detectors $a$ and $b$ intrinsically by directly predicting the corrected time difference value $\Delta t_m^{corr}$. In this regard, these models provide only implicit corrections to the timestamps, which is why we refer to them as implicit correction models.
Following this approach, it becomes apparent that a certain number of source positions is needed to generate a sufficient label variability (or class cardinality for classifier problems), otherwise it will lead to poor generalization, as shown in \cite{naunheim_holistic_2024}.\newline
To remove the need for measuring multiple source positions along the $z$-axis in order to have sufficient label variability, we propose to redefine the residuals in a way which significantly enlarges the label space even when using only one source position at the $z$-axis. This can be achieved by using the correction distribution $\{r_i \}$, where each correction $r_i$ is defined as
\begin{align}
l_i \coloneq r_i (\Delta t_{m,i}, z_i) &= \frac{\Delta t_{m,i} - \Delta t_{\mathbb{E}} (z_i)}{2}\\
\label{eq:explicitLabel}
&=\frac{\Delta t_{m,i} + \frac{2}{c} \cdot z_i}{2},
\end{align}
with $\Delta t_{m,i}$ denoting the $i$-th measured time difference and $\Delta t_{\mathbb{E}}$ the expected time difference defined in \cref{eq:implicitLabel}. From \cref{eq:explicitLabel} it becomes clear, that models trained on these labels are able to provide explicit corrections, namely that the corrected timestamps are given as
\begin{equation}
\begin{aligned}
t_{a,i}^{\text{corr}} &= t_{a,i} - r_i, \\
t_{b,i}^{\text{corr}} &= t_{b,i} + r_i.
\end{aligned}
\end{equation}
Furthermore, this explicit correction formulation (see \cref{eq:explicitLabel}) introduces a translational symmetry into the labeling process. Therefore, it does not pose any restriction on the number of sources along the $z$-axis such that the experimenter can decide if it is wanted to include a spatial sampling along $z$ during the acquisition process or if a few or even one source position is sufficient regarding practicability. If one source position should be used, the experimenter can choose an arbitrary location along the $z$-axis. However, for simplicity, we recommend setting the source to the center position since there, the expected time difference $\Delta t_{\mathbb{E}}$ equals zero, which simplifies \cref{eq:explicitLabel}. Mathematical considerations about the explicit correction formulation can be found in the appendix.\newline
When comparing both label definitions (see \cref{fig:LabelScheme}), one can see that the label distribution is discrete for the implicit approach. Precisely, the number of unique labels (unique expected time differences) is equal to the number of used source positions, which might result a sparse coverage of the label domain. Contrary to this, the explicit correction approach generates a continuous spectrum of labels, which is independent from the number of measured source positions due to translational symmetry of the formulation resulting in a high coverade of the label domain. The various label approaches also lead to differences in terms of label balance or imbalance. In the implicit approach, it is very easy to generate a high balance of labels by using the same measurement time per source position. This promotes that all labels can be trained equally well. In comparison, the explicit approach leads to a label imbalance independent of the selected measurement time. This is because the labels are based, among other things, on the measured time difference spectrum, which is Gaussian-like distributed. As a result, corrections with a large magnitude are less often represented in the training data set than corrections with a smaller magnitude.

\subsection{Evaluation Metrics}

In recent works, we proposed to use a three-fold evaluation scheme \cite{naunheim_first_2023,naunheim_holistic_2024, naunheim_using_2024} for any kind of machine learning-based \ac{TOF} correction or calibration. The scheme consists of a data scientific, physics-based and \ac{PET}-based evaluation approach. Only models that pass the first two evaluation criteria are evaluated in a subsequent step regarding their application in \ac{PET}.

\subsubsection{Data Scientific Evaluation}
The data scientific evaluation approach assesses the models based on typical metrics known from machine learning. In this work, we calculate the \ac{MAE} for all distinct $N$ label classes $\{l_1, \dots, l_{N}\}$ separately,
\begin{equation}
\text{MAE}_{k} = \frac{1}{n_k} \sum \limits_{i=1}^{n_k} ||p_{k,i} - l_{k,i} ||_1,
\end{equation}
with $n_{k}$ denoting the number of test sampled for the label class $l_k$, and $p_{k,i}$ describing the prediction associated to the $i$-th sample with ground truth $l_k$, respectively. Although the \ac{MAE} provides a numerical value, we use it rather as a qualitative metric to check if a model's performance is stable for large portions of the test data. In the first part of this work, the progression of the \ac{MAE} is analyzed. Models demonstrating an \ac{MAE} curve progression that is noticeably higher than the progression curve of the other corresponding models, or that exhibit oscillations in the \ac{MAE} progression that align with the source positions utilized during training, are considered to have failed the data scientific quality check. The second part relies on the mean \ac{MAE} and weighted mean \ac{MAE} for visualization and comparability reasons. The mean \ac{MAE} is defined by the sample mean over the $N$ label classes, 
\begin{equation}
\text{mean MAE} = \frac{1}{N} \sum \limits_{k=1}^{N} \text{MAE}_{k}.
\end{equation}
The weighted mean \ac{MAE} takes the label distribution into account by weighting the \ac{MAE} with the number of occurrences $n_k$ of the label class $l_k$,
\begin{equation}
\text{weighted mean MAE} = \left( \sum \limits_{k=1}^N n_k  \right)^{-1}  \left( \sum \limits_{k=1}^N \text{MAE}_k \cdot n_k \right)
\end{equation}
We perform the data scientific evaluation without posing any restrictions on the estimated energy of the input data.

\subsubsection{Physics-based Evaluation}
\label{subsubsec:physicsEval}
The next evaluation part considers aspects from physics, which are the linearity and a so-called $\varepsilon$-value indicating the agreement with fundamental quantities like the speed of light. For this test, data stemming from different source positions along the $z$-axis are fed into the model. The directly (implicit) or indirectly (explicit) predicted time differences are histogrammed for each source position, and the means of the corresponding prediction distributions are estimated by fitting \cite{noauthor_orthogonal_nodate} over a $3\sigma$-range a Gaussian function assuming Poissonian uncertainties on the time differences. The obtained means $\{\mu_i\}$ and the corresponding source positions $\{z_i\}$ are finally used to perform a linear regression using
\begin{equation}
\Delta t_{\mathbb{E}} \approx \mu (z; \varepsilon, b) = - \frac{2}{c} \cdot \varepsilon \cdot z + b,
\label{eq:linReg}
\end{equation}
where we assume that the mean $\mu$ matches in good approximation the expected time difference $\Delta t_{\mathbb{E}}$. For the linear regression, the uncertainty $\sigma_z$ on the source position $z$ is dominated by the active source diameter and estimated by assuming a uniform distribution of activity within this diameter ($\sigma_z = \qty{0.5}{\milli \metre}/\sqrt{12}$). In a similar fashion, the uncertainty $\sigma_{\mu}$ on the estimation of mean time difference $\mu$ is given by assuming an uniform distribution within the histogram bin width ($\sigma_{\mu} = \qty{10}{\pico \second}/\sqrt{12}$).\newline
\Cref{eq:linReg} closely resembles the fundamental \cref{eq:implicitLabel}, with both describing a linear dependence between the expected time difference and the source position. We evaluate qualitatively if this assumption is fulfilled by the predictions of our models using reduced $\chi^2$-values, which should be in the optimal case closely distributed around one. For the sake of easier comparability of many models, in the second part of this work, we compress the $\chi^2$-distribution information into a single value $s_{\chi^2}$ measuring the spread of the $\chi^2$-value from the optimal value of 1 in units of the standard deviation $\sigma_{\chi^2}$,
\begin{equation}
s_{\chi^2} = \frac{|\overline{\chi^2} - 1|}{\sigma_{\chi^2}},
\end{equation}
with $\overline{\chi^2}$ denoting the mean.\newline
As a quantitative measure, we use the number of $\sigma$-environments $n_{\overline{\epsilon}}$ of the $\varepsilon$-value minding the theory value of $\varepsilon_{\text{theo}} = 1$,
\begin{equation}
n_{\overline{\varepsilon}} = \frac{|\overline{\varepsilon} - \varepsilon_{\text{theo}}|}{\sigma_{\overline{\varepsilon}}}.
\end{equation}
In order to suppress any deteriorating effects coming from time walk, events are selected to be in an energy window from \qtyrange{430}{590}{\kilo \eV} for the physics-based evaluation.
A model passes the physics-based evaluation, if the $\chi^2$-values are closely distributed around 1, or the spread $s_{\chi^2}$ in units of the standard deviation is smaller than 3. Furthermore, the models must demonstrate their agreement with physics by having $\overline{\varepsilon}$-values being compatible with the theory in a $3\sigma$-range.

\subsubsection{PET-based Evaluation}
If the data scientific and physics-based evaluations are positive, finally, the performance of the model is tested assessing the achievable \ac{CTR}. For this, data is fed into the models, and the resulting predictions are filled into a histogram. The \ac{FWHM} of this distribution is estimated by fitting a Gaussian function assuming Poissonian uncertainties on the time differences. We perform this analysis for test data from the iso-center (describing the center of the setup ($x=0, y=0, z=0$) \unit{\milli \metre}) but also for test data along the complete $z$-axis. 

\subsection{Model Architecture}

We use \acp{GBDT} as model architecture for the second part of the residual physics-based timing calibration. \acp{GBDT} are a classical machine learning approach with the advantage that they can handle missing values and allow the implementation in a \ac{FPGA} \cite{krueger_high-throughput_2023}. This makes it especially suitable for the application in \ac{PET} scanners minding edge-\ac{AI} approaches \cite{zhou_edge_2019, deng_edge_2020}. The model is based on an ensemble of binary decision trees, which uses boosting as a learning procedure. Each tree that is added to the ensemble during the training process attempts to minimize the errors of the predecessor model. In this study, we worked with the implementation of XGBoost \cite{chen_xgboost_2016}, which has proven it's predictive power \cite{grinsztajn_why_2024} in our prior studies \cite{naunheim_improving_2023,naunheim_holistic_2024}.\newline
Like in many other machine learning algorithms, several hyperparameters can be set before training a model. The maximum tree depth $d$ describes the maximum number of decisions inside a single tree. The learning rate $lr$ works as a smoothing parameter and controls the contribution of each new tree to the overall model. Another hyperparameter is the maximum number of trees within the ensemble, which is often used in combination with an early stopping criterion that stops the training procedure as soon as the validation loss has not improved for a certain amount of boosting rounds. The \ac{MR} \cite{muller_gradient_2018} of a single decision tree is given to be
\begin{equation}
\text{MR}(d) =  \left(2^d - 1 \right) \cdot \qty{11}{\byte} + 2^d \cdot \qty{6}{\byte}.
\label{eq:MR}
\end{equation}

\subsection{Study Design}
\label{subsec:StudyDesign}

Several datasets were recorded in order to realize the different study designs for evaluating the implicit and explicit correction approach. In the following text, we will often refer to the coordinate system defined in \cref{fig:LabelScheme}.\newline

\begin{figure}[ht]
\centering
\includegraphics[height=7cm]{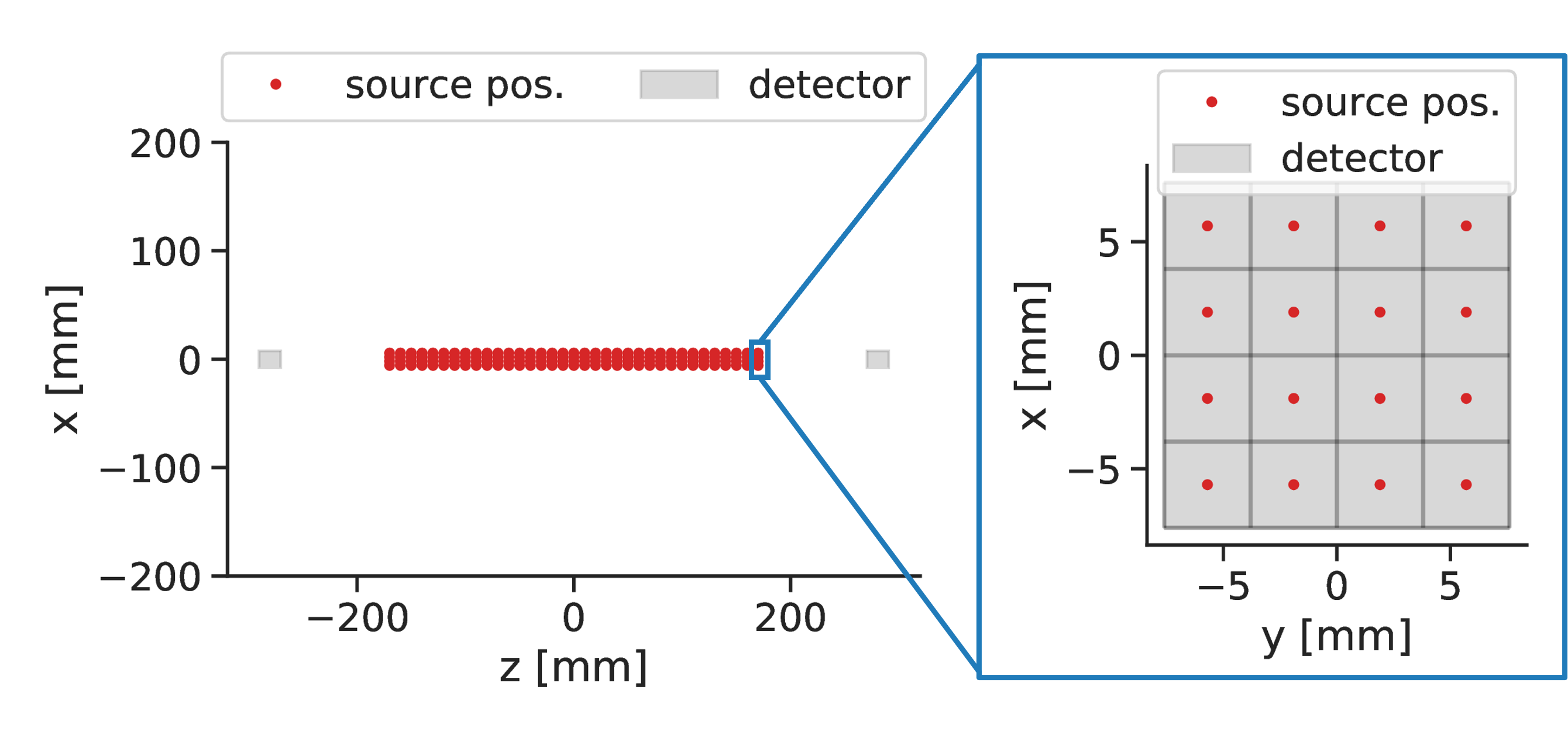}
\caption{Visualization of the source distribution for a part of the transaxial study.}
\label{fig:comp_pattern}
\end{figure}

\subsubsection{The Transaxial Performance Study}
\label{subsubsec:tps}

\begin{figure}[ht]
\centering
\includegraphics[width=\scBroadd \textwidth]{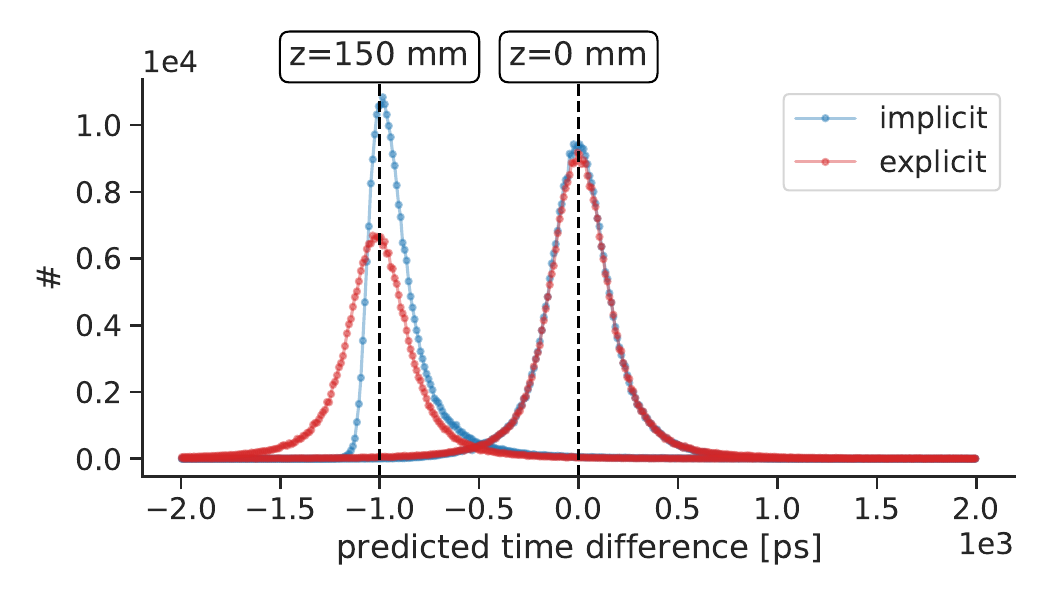}
\caption{Examples of the prediction distribution of implicit and explicit correction models. While the explicit correction model preserves the Gaussian time difference shape, the implicit correction model is strongly affected by edge effects leading to non-Gaussian distributions. In this example, the implicit model IM$_{10,12}$ and the explicit model EM$_{10,4}$ is used.}
\label{fig:example_pred}
\end{figure}

This study aims to compare both calibration approaches regarding their performance along the axis connecting both detector stacks. To acquire data for training, validation and testing \num{35} positions along the $z$-axis were utilized, while at each $z$-position, a grid of \numproduct{4x4} in-plane positions are measured. The in-plane positions were chosen so that each crystal segmented was centrally irradiated. The step width (sw) between two subsequent planes was set to sw=\qty{10}{\milli \metre}, with the maximum $z$-positions given to be $z_{\text{min/max}} = \pm \qty{170}{\milli \metre}$, which equals the maximum travel distance allowed by the translation stage system. The distribution of the source positions is shown in \cref{fig:comp_pattern}. For each source position a measurement time of \qty{45}{\minute} was used.\newline
From the acquired dataset, \num[round-mode=places, round-precision=2, exponent-mode=scientific]{26391009} and \num[round-mode=places, round-precision=2, exponent-mode=scientific]{8797003} samples are used in total to form the training and validation dataset (66/33 split), respectively. The numbers of samples per $z$-position show a deviation $<\qty{2}{\percent}$, such that bias effects due to different solid angles can be ruled out. The remaining data of the measurement run, namely \num[round-mode=places, round-precision=2, exponent-mode=scientific]{105252579} samples, is used to build the test dataset.\newline
In addition to the experiment, where each $z$-position is present in the training data, we removed data along the $z$-axis to artificially create new training and validation datasets with different step widths (sw=\qty{50}{\milli \metre} and sw=\qty{100}{\milli \metre}) between the grid planes. Although this results in a lower number of training samples, we follow this approach to analyze how implicit and explicit correction models perform at unknown $z$-position. Therefore, new models are trained on these datasets but tested on the finely sampled and unseen test dataset comprising \num{35} $z$-positions. \newline
For both kinds of training approaches (implicit and explicit), the same amount of training and validation samples are used, and a big parameter space of the maximum tree depth $d$ is sampled, ranging from very narrow to very deep trees ($d \in \{4, 8, 12, 16, 20\}$). The learning rate is always set to $lr=0.1$, the maximum number of trees in an ensemble is set to \num{1000}, and we use \num{10} early stopping rounds.\newline
All models are evaluated regarding their \ac{MAE} performance and linearity behavior.
In prior studies \cite{naunheim_improving_2023,naunheim_holistic_2024}, we demonstrated that this results in edge effects leading to non-Gaussian distribution (see \cref{fig:example_pred}). For this reason, the linearity analysis of the implicit correction models is performed for a large central region (\qtyrange{-100}{100}{\milli \metre}), while the explicit correction approach allows an analysis over the complete $z$-range (\qtyrange{-170}{170}{\milli \metre}). Models that pass the data scientific and physics-based quality control are finally evaluated regarding their \ac{CTR} improvement on input data from one $z$-position ($z=\qty{0}{\milli \metre}$). We denote an implicit model that has been trained with data utilizing a step width of sw and a maximum tree depth of d as IM$_{(\text{sw},\text{d})}$, while the corresponding explicit model is called EM$_{(\text{sw},\text{d})}$.

\subsubsection{The In-Plane Distribution Study}

\begin{figure}[htb]
\centering
\subfloat[\numproduct{1x1} grid.]{%
    \includegraphics[width=0.3\textwidth]{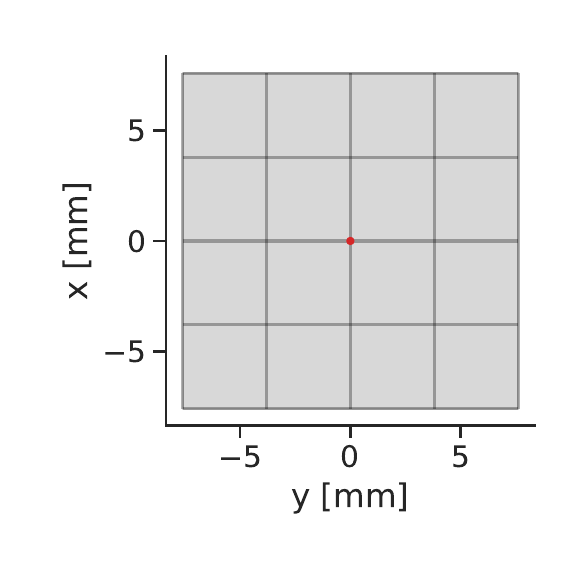}
    \label{fig:subfig1}
}
\hfill
\subfloat[\numproduct{3x3} grid.]{%
    \includegraphics[width=0.3\textwidth]{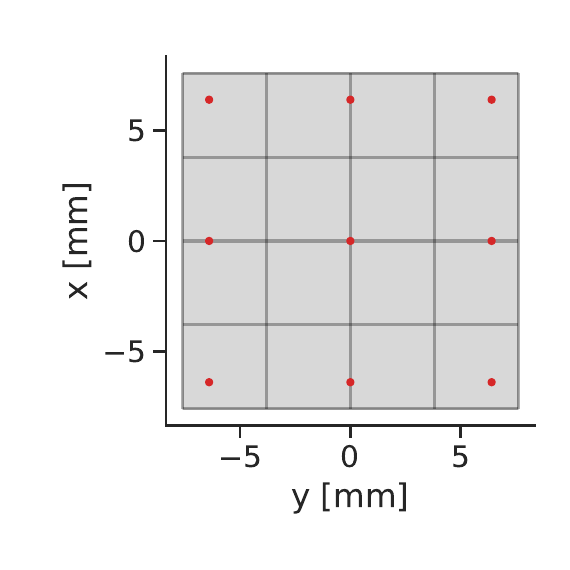}
    \label{fig:subfig2}
}
\hfill
\subfloat[\numproduct{5x5} grid.]{%
    \includegraphics[width=0.3\textwidth]{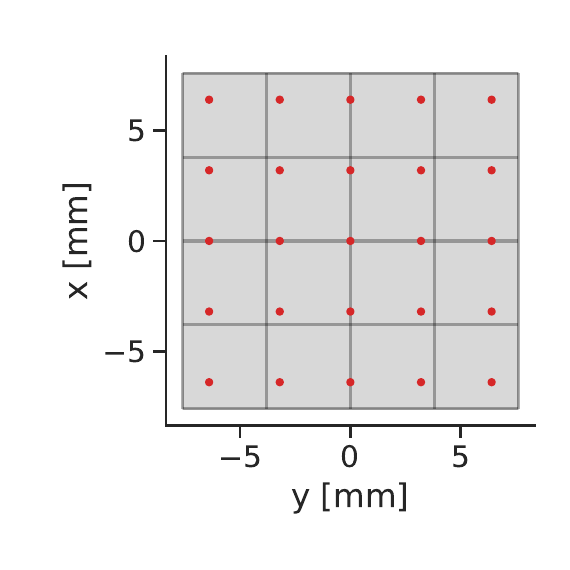}
    \label{fig:subfig3}
}

\vspace{0.5cm}

\subfloat[\numproduct{9x9} grid.]{%
    \includegraphics[width=0.3\textwidth]{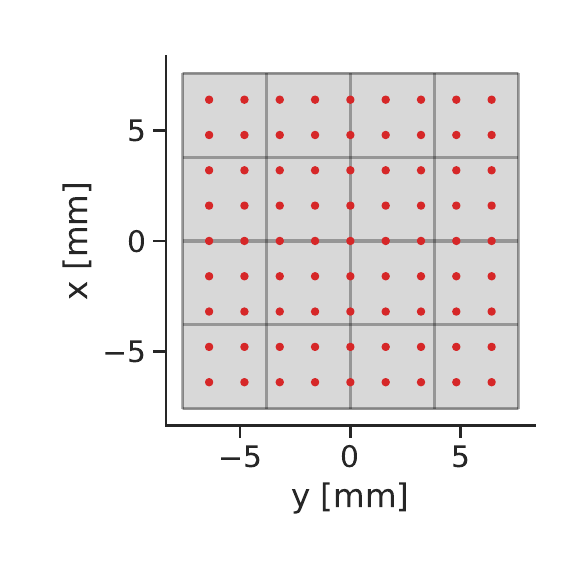}
    \label{fig:subfig4}
}
\hfill
\subfloat[\numproduct{17x17} grid.]{%
    \includegraphics[width=0.3\textwidth]{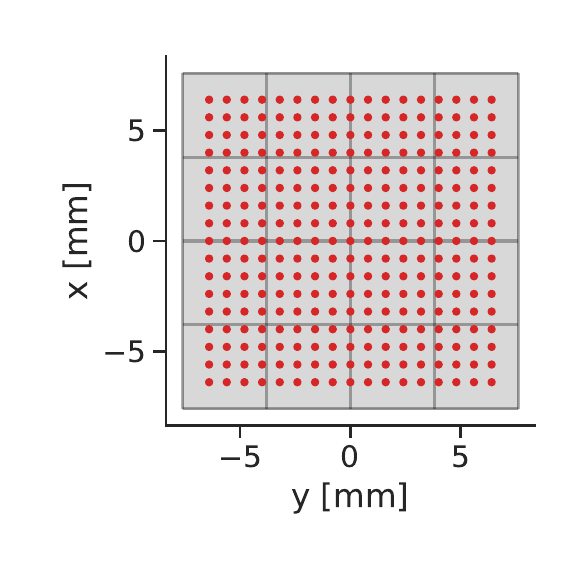}
    \label{fig:subfig5}
}
\hfill
\subfloat[\numproduct{33x33} grid.]{%
    \includegraphics[width=0.3\textwidth]{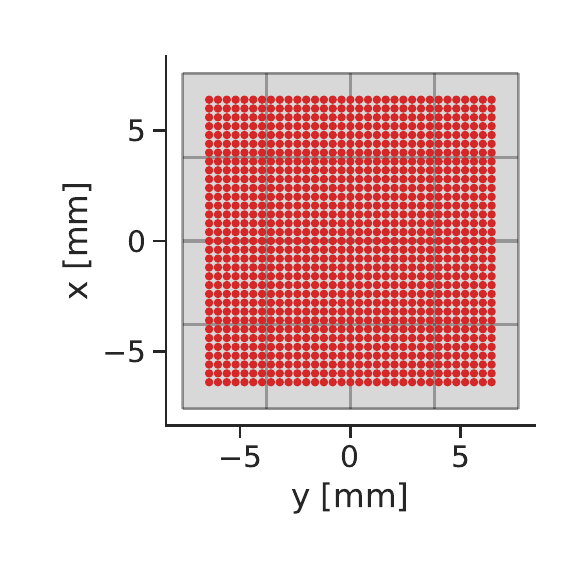}
    \label{fig:subfig6}
}

\caption{In-plane source distributions. The red dots represent the position of a source, while the gray background displays the detector.}
\label{fig:InPlaneSourceDist}
\end{figure}

This study aims to evaluate how the in-plane source distribution affects the performance of explicit correction models. Data is acquired at only one $z$-position ($z=\qty{0}{\milli \metre}$), but with an extended source distribution of \numproduct{33 x 33} source positions in the $x$-$y$-plane. At each source position a measurement time of \qty{5}{\minute} is to obtain the same number of training and validation samples as in the transaxial performance study. Six distinct new datasets for the training of explicit correction models are formed from the data, by removing sources in a regular pattern until only one source located at the iso-center remains. The in-plane distributions (IPD) are quadratic
\begin{equation}
\text{IPD} \in \{ (\numproduct{1x1}), (\numproduct{3x3}), (\numproduct{5x5}), (\numproduct{9x9}), (\numproduct{17x17}), (\numproduct{33x33})\},
\end{equation}
which also can be seen \cref{fig:InPlaneSourceDist}. Identical to the transaxial performance study, five explicit correction models were trained for each dataset with the maximum tree depth $d$ being $d \in \{4, 8, 12, 16, 20\}$. The models are tested on the transaxial test dataset comprising 35 $z$-positions. The same evaluation scheme like in the transaxial performance study is used, which incorporates data scientific, physics-based and PET-based metrics.

\subsection{Data Pre-Processing}

Hits corresponding to a single $\gamma$-interaction must be grouped into clusters by analyzing their initial timestamps. This clustering approach enables the aggregation of all relevant information linked to a specific $\gamma$-interaction. Following this step, clusters are further grouped into coincidences by applying a coincidence window, which considers the time differences between two clusters using their main \ac{SiPM}. Hits with energy values (in arbitrary units) below \num{2.5} or above \num{100} were excluded from the clustering process. A cluster window of \qty{8}{\nano \second} was empirically chosen by analysis of the temporal distribution. Furthermore, a coincidence window of \qty{50}{\nano \second} was applied to significantly reduce the proportion of random events and minimize the risk of excluding true coincidences.\newline
To each cluster, a $\gamma$-interaction position and timestamp value was assigned based on the coordinates and timestamp of the SiPM with the highest energy value. The energy signals were saturation corrected using the \qty{511}{\kilo \eV} and \qty{1275}{\kilo \eV} energy peak of the $^{22}$Na source. An energy value derived from the deposited energy on the main SiPM was assigned to each cluster.\newline
Following our proposed residual physics-based calibration scheme, a conventional time skew calibration \cite{naunheim_analysis_2022} is conducted in order to remove the major skews of the first order.

\subsection{Input Features}
The input features for both models are derived from the detection information of coincident clusters and can be separated into a set with information from detector $a$, and a set with information about detector $b$. Since both sets are similar content-wise, we describe in a general way the information a sample is made up of.\newline
The $n_{a/b}$ first relative timestamp values, which have been corrected using the first-order analytical calibration, along with their \ac{SiPM} ID, represent the temporal information. Furthermore, we use the magnitude of the detected energy signal on these \acp{SiPM}, and the ID, energy signal, and coordinates of the main \ac{SiPM} (\ac{SiPM} detecting the highest number of photons) as additional information. In addition to this, the row and column energy sums are used. Furthermore, the number of triggered \acp{SiPM}, and the difference between the first and last timestamp of a cluster are fed to the model. Finally, the first (\ac{COG}) and second moment of the light distribution are also included. Missing values are handled as NaN.\newline
While these quantities represent the input for the explicit correction models, the implicit correction models use one additional feature: the measured time difference. Based on a prior statistical analysis on how many triggered \acp{SiPM} are contained in a cluster, $n_{a/b}$ is set to twelve and nine. No filter is applied to the energy or light distribution of the input data to ensure that the trained models function effectively with any potential input sample.

\begin{figure}[hb]
\centering
\subfloat[Implicit correction models.]{%
    \includegraphics[width=0.45\textwidth]{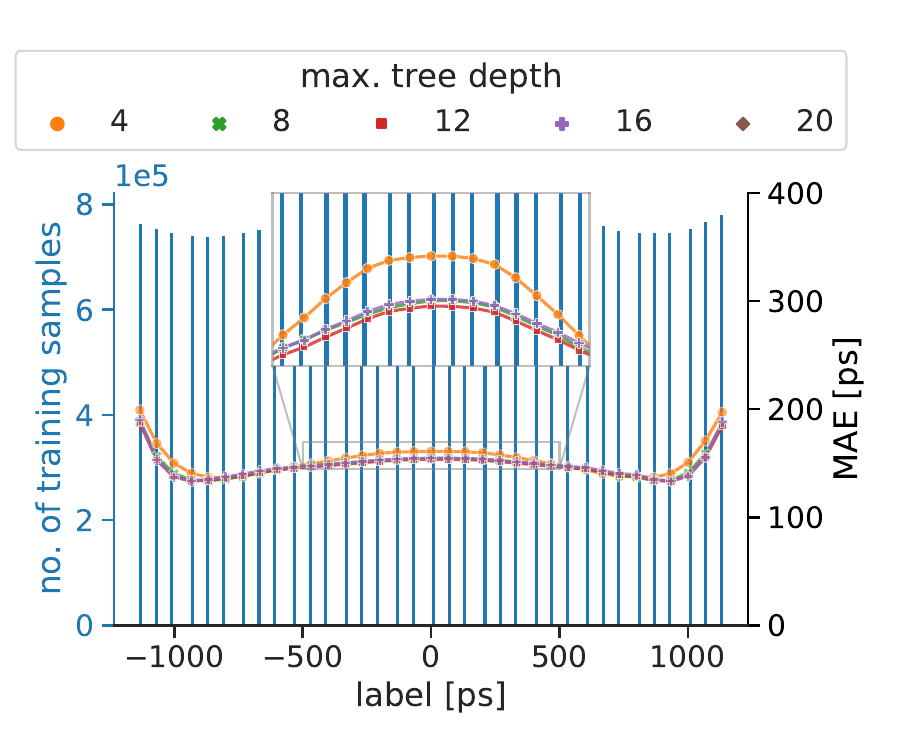}
    \label{fig:mae10_i}
}
\hfill
\subfloat[Explicit correction models.]{%
    \includegraphics[width=0.45\textwidth]{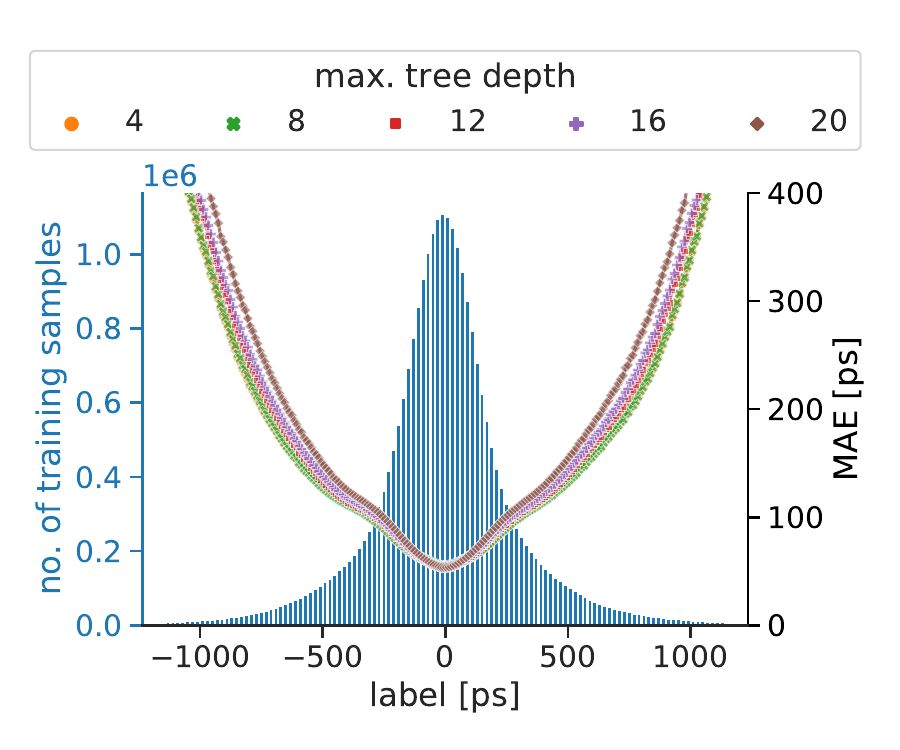}
    \label{fig:mae10_e}
}

\caption{\ac{MAE} progression of the correction models trained on a dataset using a step width of \qty{10}{\milli \metre}, and tested on unseen data with a step width of \qty{10}{\milli \metre}. The blue axis and histogram display the label distribution of the training dataset. The \ac{MAE} progression IM$_{10, 20}$ cannot be seen in plot (a), since it resulted in a very high \ac{MAE} of about \qty{1000}{\pico \second}, being far outside the displayed range.}
\label{fig:study1_mae10}
\end{figure}

\section{Results}
\subsection{The Transaxial Performance Study}
\subsubsection{Data Scientific Evaluation}

\begin{figure}[tb]
\centering
\subfloat[Implicit correction models.]{%
    \includegraphics[width=0.45\textwidth]{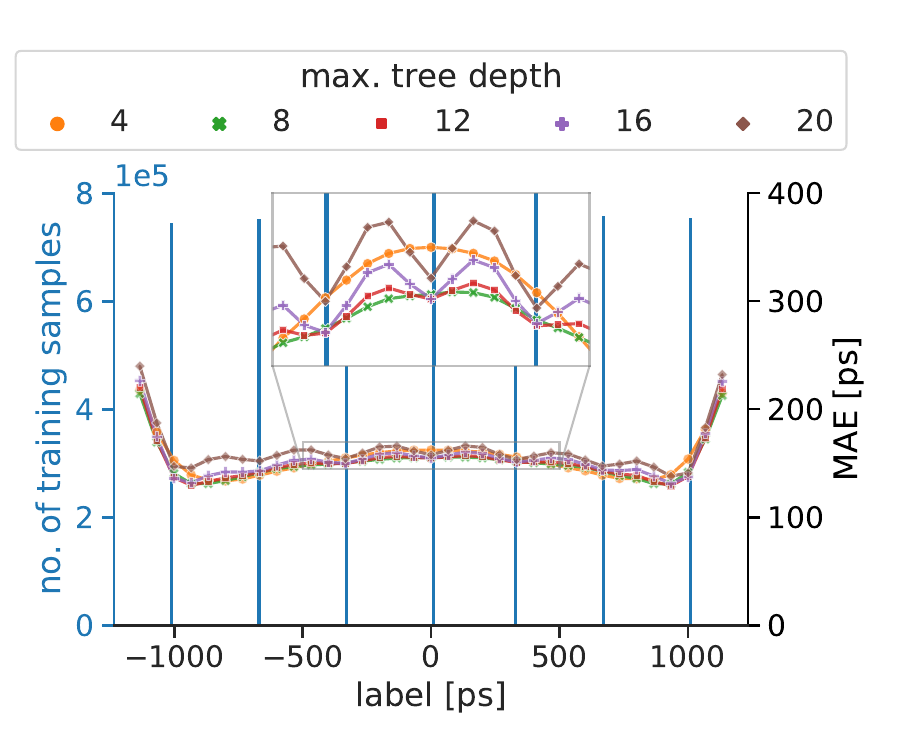}
    \label{fig:mae50_i}
}
\hfill
\subfloat[Explicit correction models.]{%
    \includegraphics[width=0.45\textwidth]{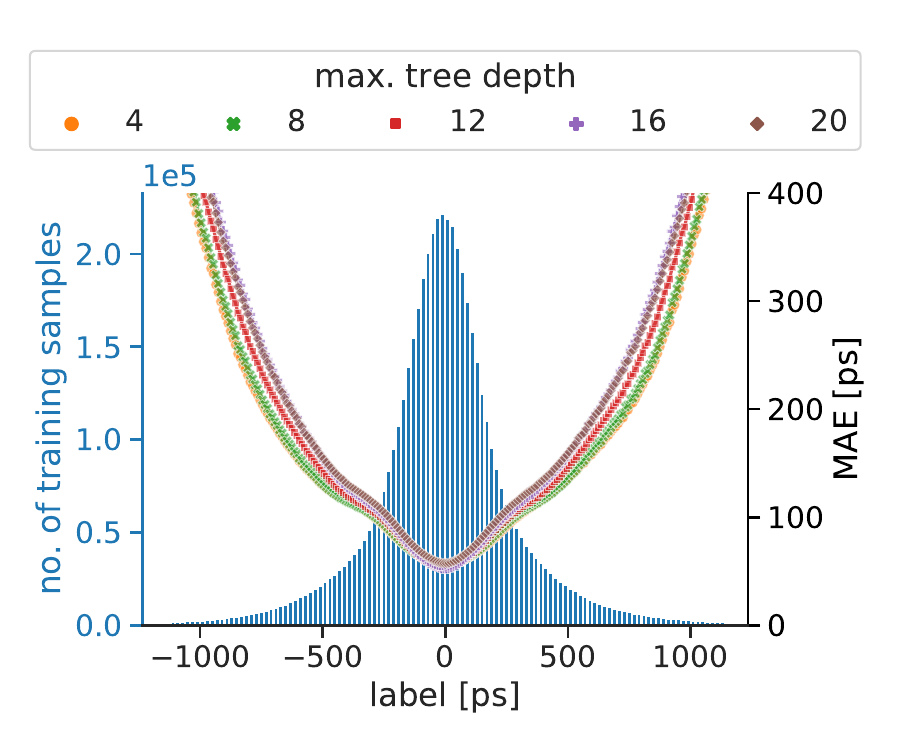}
    \label{fig:mae50_e}
}

\caption{\ac{MAE} progression of the correction models trained on a dataset using a step width of \qty{50}{\milli \metre}, and tested on unseen data with a step width of \qty{10}{\milli \metre}. The blue axis and histogram display the label distribution of the training dataset.}
\label{fig:study1_mae50}
\end{figure}

\begin{figure}[htb]
\centering
\subfloat[Implicit correction models.]{%
    \includegraphics[width=0.45\textwidth]{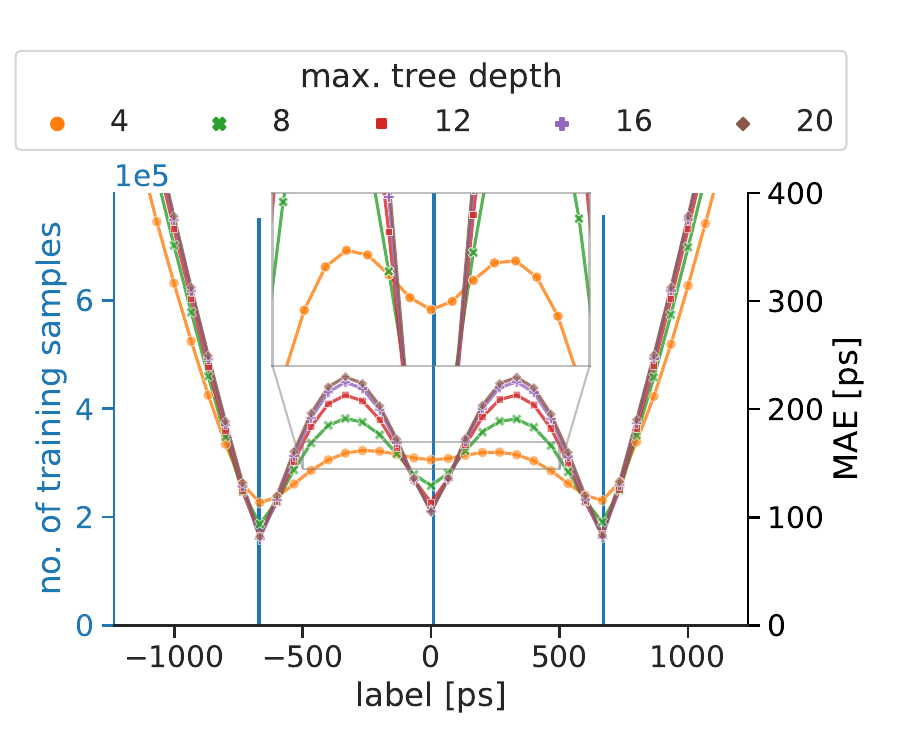}
    \label{fig:mae100_i}
}
\hfill
\subfloat[Explicit correction models.]{%
    \includegraphics[width=0.45\textwidth]{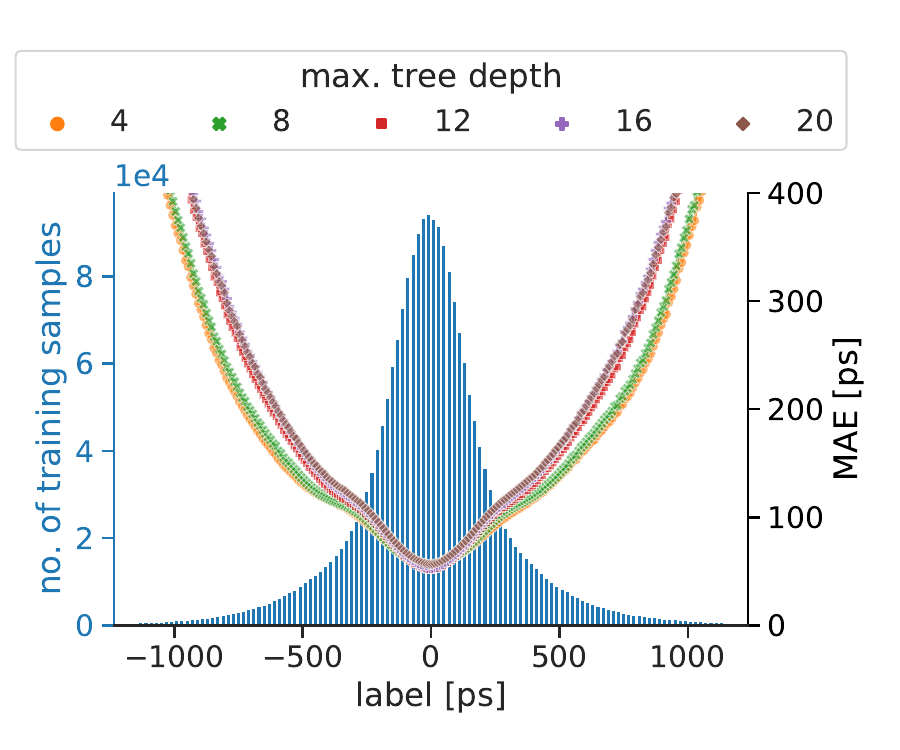}
    \label{fig:mae100_e}
}

\caption{\ac{MAE} progression of the correction models trained on a dataset using a step width of \qty{100}{\milli \metre}, and tested on unseen data with a step width of \qty{10}{\milli \metre}. The blue axis and histogram display the label distribution of the training dataset.}
\label{fig:study1_mae100}
\end{figure}

Prior to assessing the models with the three-fold evaluation scheme, we checked that the models were trained successfully by analyzing the training and validation curves. All models showed a smoothing training progression and were fully trained out.\newline
The \ac{MAE} progression of the explicit and implicit correction models for different training step width is displayed in \cref{fig:study1_mae10}, \cref{fig:study1_mae50}, and \cref{fig:study1_mae100}. For the models trained on data with a step width of \qty{10}{\milli \metre}, both correction models show a smooth \ac{MAE} progression without any outliers. While most of the \ac{MAE} values of the implicit correction models are located in a region between \qtyrange{100}{200}{\pico \second}, the \ac{MAE} rises towards the edges of the testing data. The labels of the training dataset for implicit correction models are, in good approximation, uniformly distributed. The \ac{MAE} progression of the explicit correction models resembles a U-form, with the lowest \ac{MAE} of sub-\qty{100}{\pico \second} being achieved at labels of small magnitude. When moving to higher label magnitudes, and therefore, also to higher correction magnitudes, the \ac{MAE} strongly rises. The explicit label distribution follows, in good approximation, a Gaussian shape with non-Gaussian tails.\newline
For both correction approaches trained on \qty{10}{\milli \metre} step width data, no significant performance differences for different maximum depths of the trees can be observed, except for the implicit correction model with maximum depth \num{20} providing by far the worst \ac{MAE} performance (\acp{MAE} $>$ \qty{400}{\pico \second}).

The \ac{MAE} performances of models trained on data with fewer source positions on the $z$-axis compared to the test data are shown in \cref{fig:study1_mae50} and \cref{fig:study1_mae100}. While the spatial undersampling has no big impact on the performance of the explicit models, it does have a strong effect on implicit correction models. It can be observed that for the explicit approach the label distributions between spatially full-sampled and undersampled training data does not differ. Thus, the overall \ac{MAE} progression of the models trained on \qty{50}{\milli \metre} and \qty{100}{\milli \metre} data is strongly similar to the finely sampled training data with \qty{10}{\milli \metre} stepping. It can be stated that there is a tendency for models with small maximum tree depth to provide slightly better results for spatially undersampled data.\newline

\begin{figure}[b]
\centering
\subfloat[\qty{10}{\milli \metre} stepping in training data.]{%
    \includegraphics[width=0.45\textwidth]{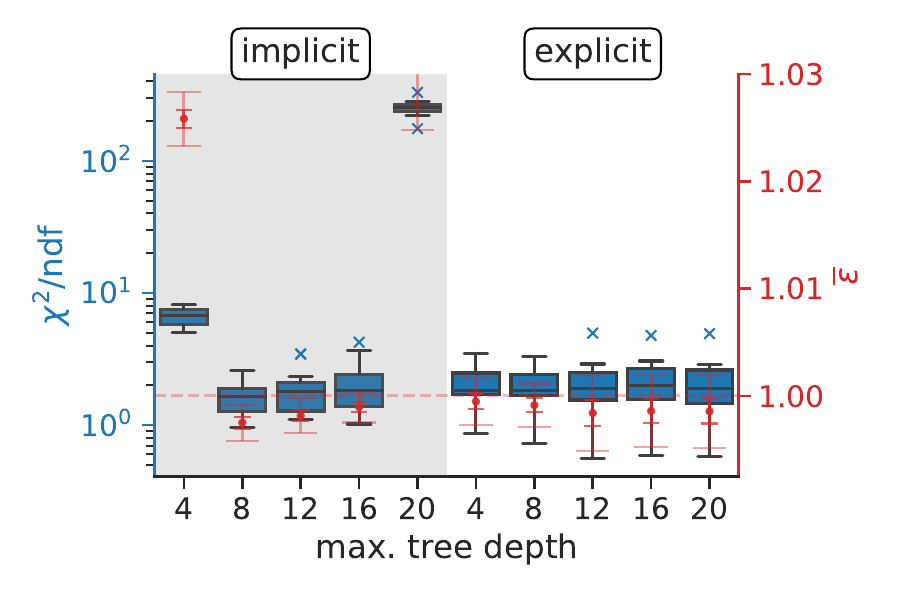}
    \label{fig:L10}
}
\hfill
\subfloat[\qty{50}{\milli \metre} stepping in training data.]{%
    \includegraphics[width=0.45\textwidth]{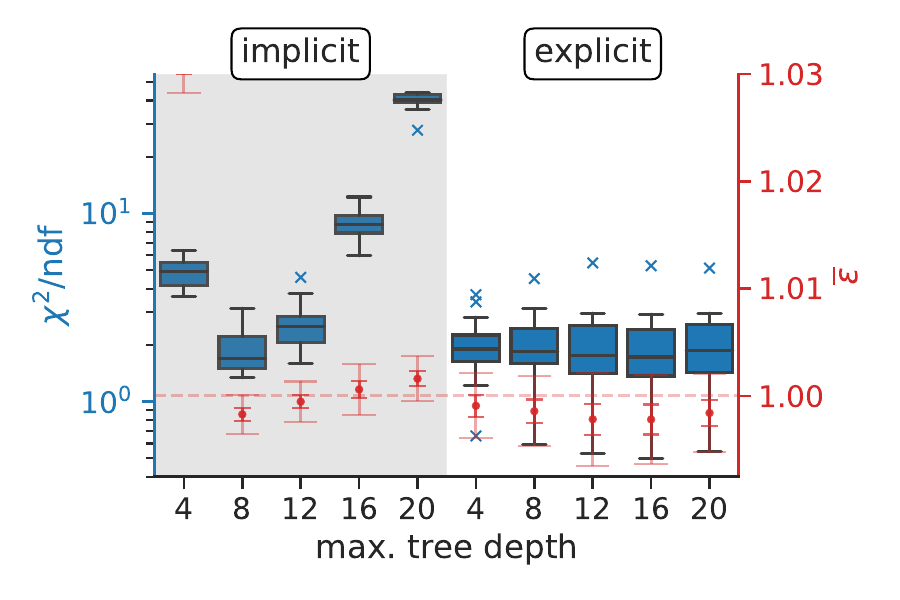}
    \label{fig:L50}
}
\vspace{0.5cm}
\subfloat[\qty{100}{\milli \metre} stepping in training data.]{%
    \includegraphics[width=0.45\textwidth]{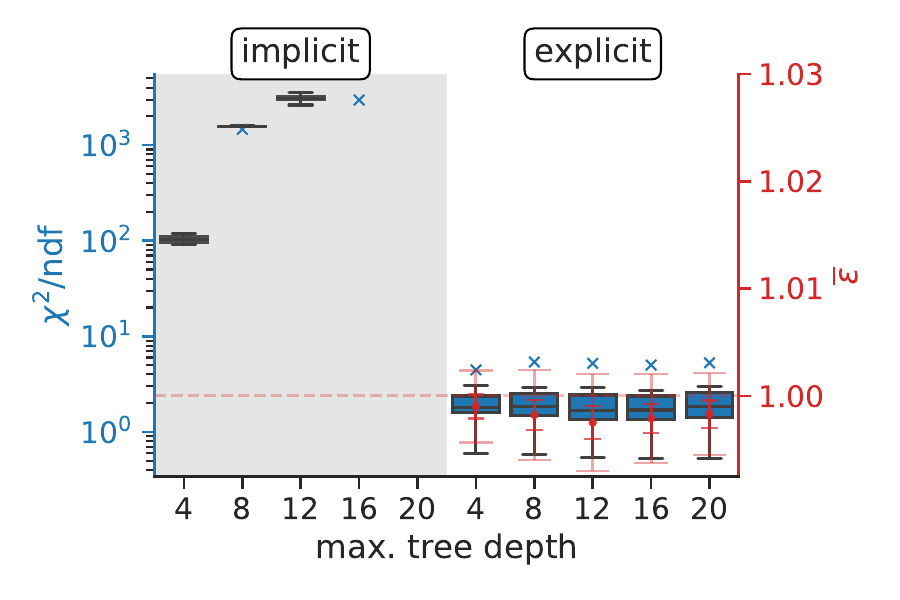}
    \label{fig:L100}
}

\caption{Linearity evaluation for the implicit and explicit models trained on training data with different step widths. The blue box plots correspond to the blue axis on the left and represent the reduced $\chi^2$-distribution. $\chi^2$/ndf values near to one indicate a good agreement with the linearity hypothesis. The red dots correspond to the axis on the right and represent the average $\varepsilon$-value derived from the linear regressions. The thick and light error bars indicate $1\sigma$ and $3\sigma$ uncertainty.}
\label{fig:L_Study}
\end{figure}

The implicit correction models are strongly affected by spatial undersampling, which can also be seen in the corresponding label distribution of the training data. Oscillation effects in the \ac{MAE} progression can be seen, and become more dominant the bigger the spatial undersampling is, which was first observed and studied in  \cite{naunheim_holistic_2024}. In addition, implicit correction models with low maximum tree depth show a more robust behavior against the spatial undersampling than models with higher complexity. This can especially be seen in \cref{fig:study1_mae50}a, where the zoom-in window clearly reveals oscillation effects for the implicit models starting from a maximum tree depth of \num{12}. Furthermore, the figure also suggests that the model with a maximum tree depth of \num{8} shows very minor oscillation behavior, being in a transition zone between non-oscillation ($d=4$) and oscillation behavior ($d=12$).\newline
The observations show that all explicit correction models meet the data scientific quality check. All implicit correction models pass the quality check for a training step width of \qty{10}{\milli \metre}, except for IM$_{10,20}$. For a stepping of \qty{50}{\milli \metre}, this reduces to the implicit correction models with a maximum tree depth of \num{4} and \num{8}, while for a step width of \qty{100}{\milli \metre} only the model with maximum tree depth \num{4} passes the evaluation, although minor oscillation tendencies are visible.

\clearpage

\subsubsection{Physics-based Evaluation}
The estimated $\chi^2$-analysis and $\varepsilon$-values are depicted in \cref{fig:L_Study}. Models that show a high compatibility with our physical expectations can be identified by $\chi^2$/ndf-values near to one, and an $\varepsilon$-value that matches within $3\sigma$ the theoretical value of $\varepsilon_{\text{theo}}=1$. The explicit correction models show good agreement with linearity and high compatibility with the theoretical $\varepsilon$-value of $\varepsilon_{\text{theo}} = 1$ across all tested step widths and tree depths. Contrary to this, there is the trend that implicit correction models lose their linearity property with increasing step width. While for the smallest step width of \qty{10}{\milli \metre}, IM$_{10,12}$ and IM$_{10, 16}$ meet the physics-based quality criteria, it remains only IM$_{50,8}$ and IM$_{50,12}$ for a training step width of \qty{50}{\milli \metre} and for the highest step width no implicit model passes the physics-based quality check.

\begin{figure}[ht]
\centering
\includegraphics[width=\scBroadd \textwidth]{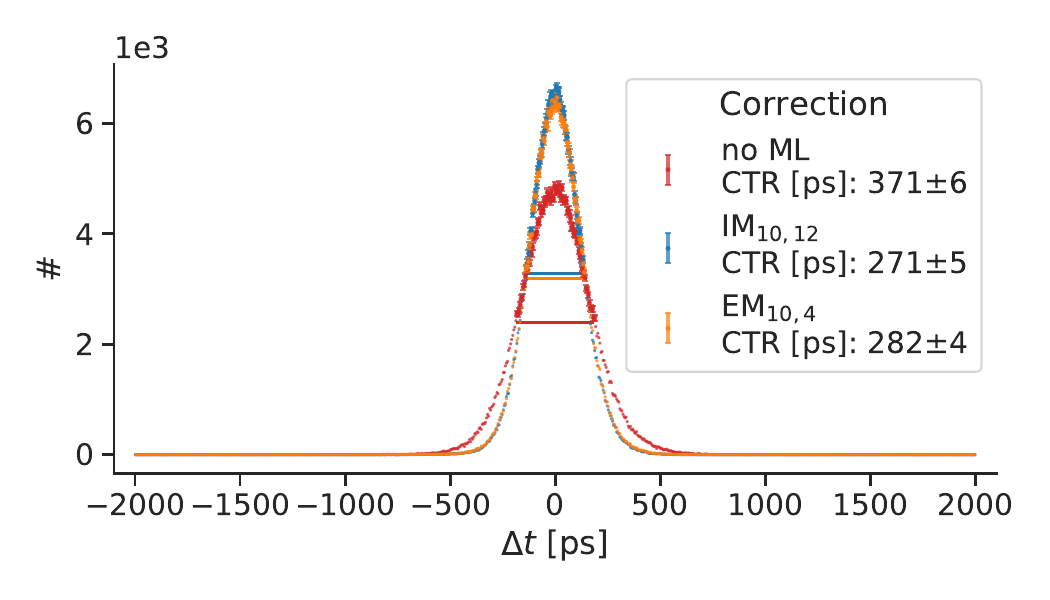}
\caption{Obtained timing resolutions for different implicit and explicit correction models. The values are based on coincidences being in an energy window from \qtyrange{430}{590}{\kilo \eV}. The correction model 'no ML' refers to performing only an analytical time skew calibration (first part of the residual physics calibration scheme), without subsequent use of machine learning.}
\label{fig:ctr_pred}
\end{figure}

\subsubsection{PET-based Evaluation}

\begin{figure}[tb]
\centering
\subfloat[Implicit correction models.]{%
    \includegraphics[width=0.45\textwidth]{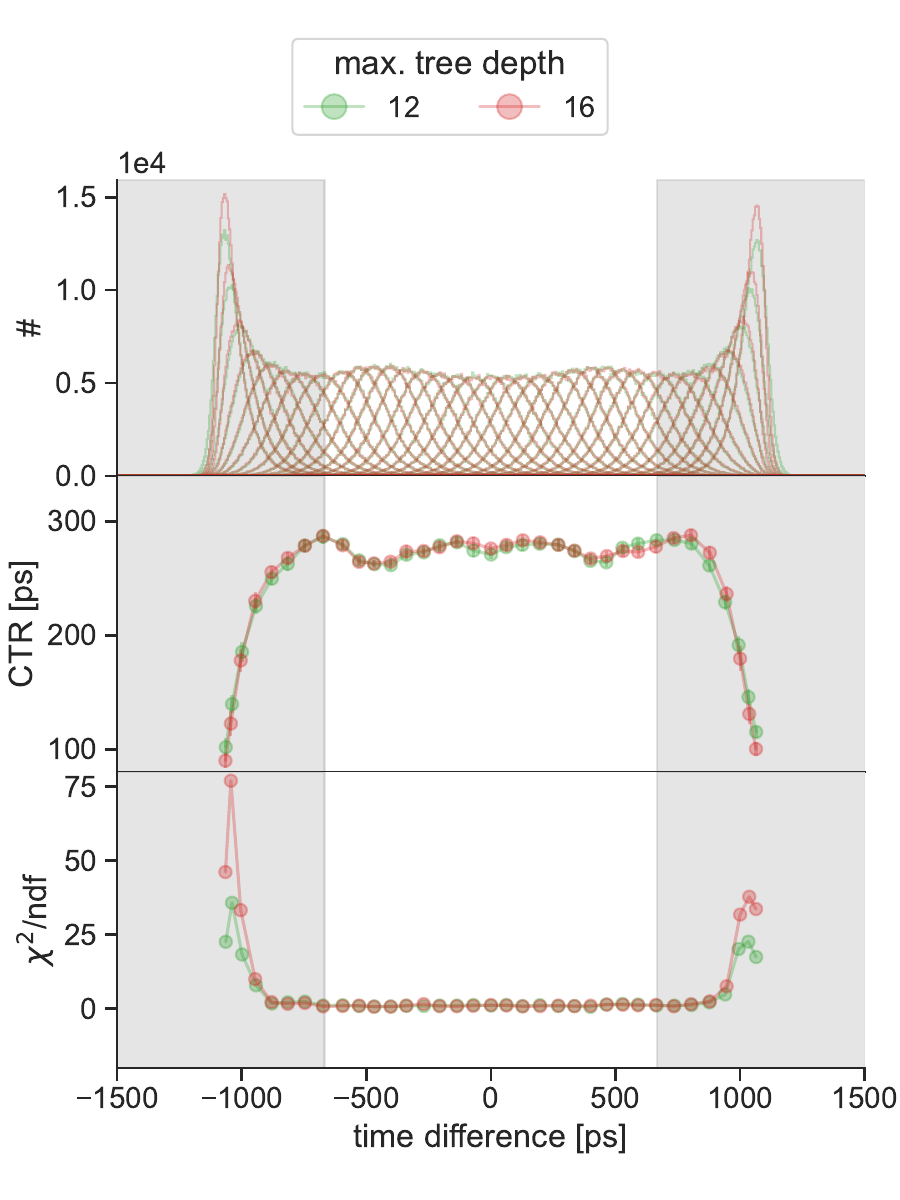}
    \label{fig:ctr_z_im}
}
\hfill
\subfloat[Explicit correction models.]{%
    \includegraphics[width=0.45\textwidth]{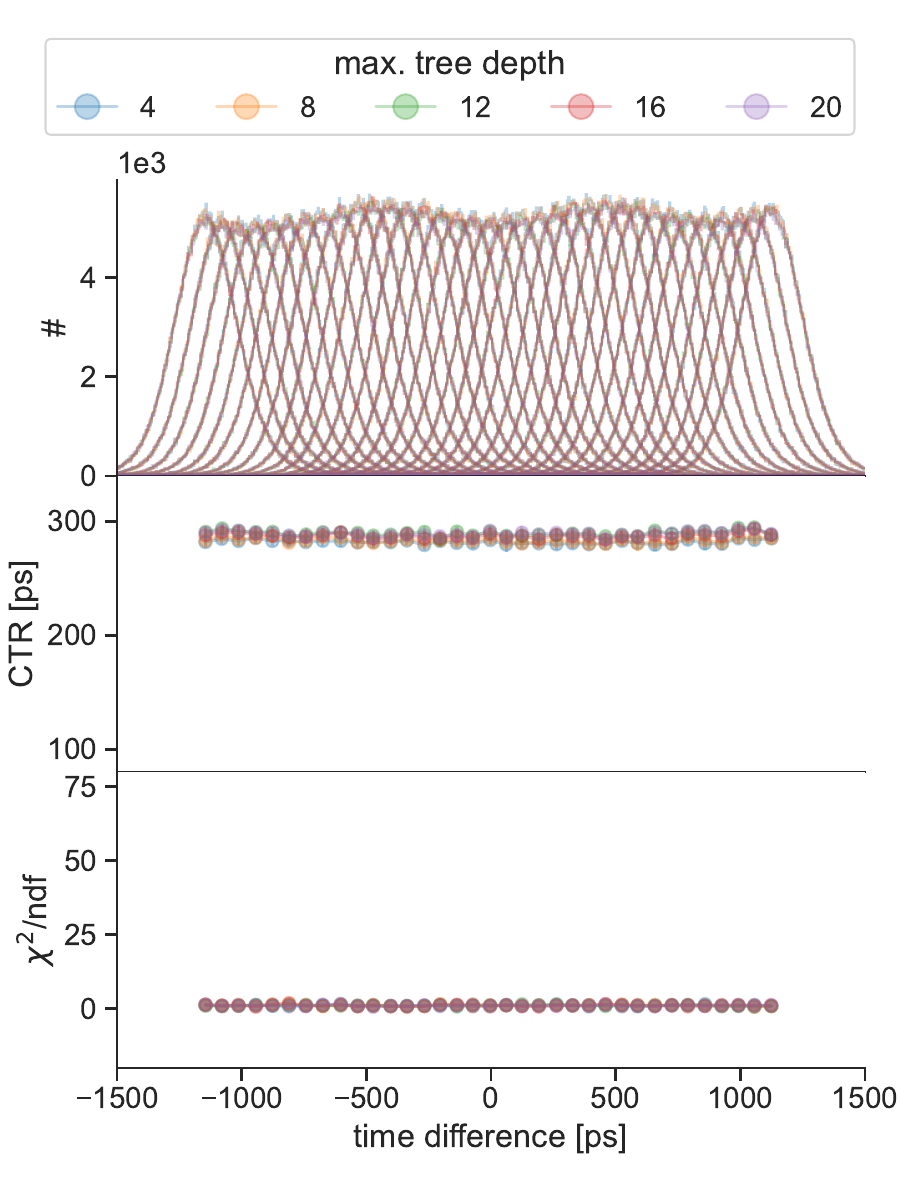}
    \label{fig:ctr_z_ex}
}

\caption{\ac{CTR} progression along the $z$-axis of models, trained on \qty{10}{\milli \metre} stepping, which passed the data scientific and physics-based quality checks. The upper plot visualizes the (directly/indirectly) predicted time differences. The middle plot shows the \ac{CTR} value at the specific position. The lower plot visualizes the $\chi^2$/ndf-value received from fitting a Gaussian function to the corresponding time difference distribution. The white area represents the source positions, that had been considered during the linearity analysis. While the explicit correction approach remains functional across the full $z$-range, the implicit correction approach is known to experience bias effects at the edges of the training domain. This leads to non-Gaussian time distributions, recognizable by the high $\chi^2$/ndf-values that return decreased \acp{CTR} values due to the distribution deformation. In order to provide an unbiased evaluation, the source positions from the gray area are excluded from the evaluation of the implicit correction approach. The data shown in this plot was selected to be within an energy window from \qtyrange{430}{590}{\kilo \eV}.}
\label{fig:ctr_z}
\end{figure}

\begin{figure}[ht]
\centering
\includegraphics[width=\scBroad \textwidth]{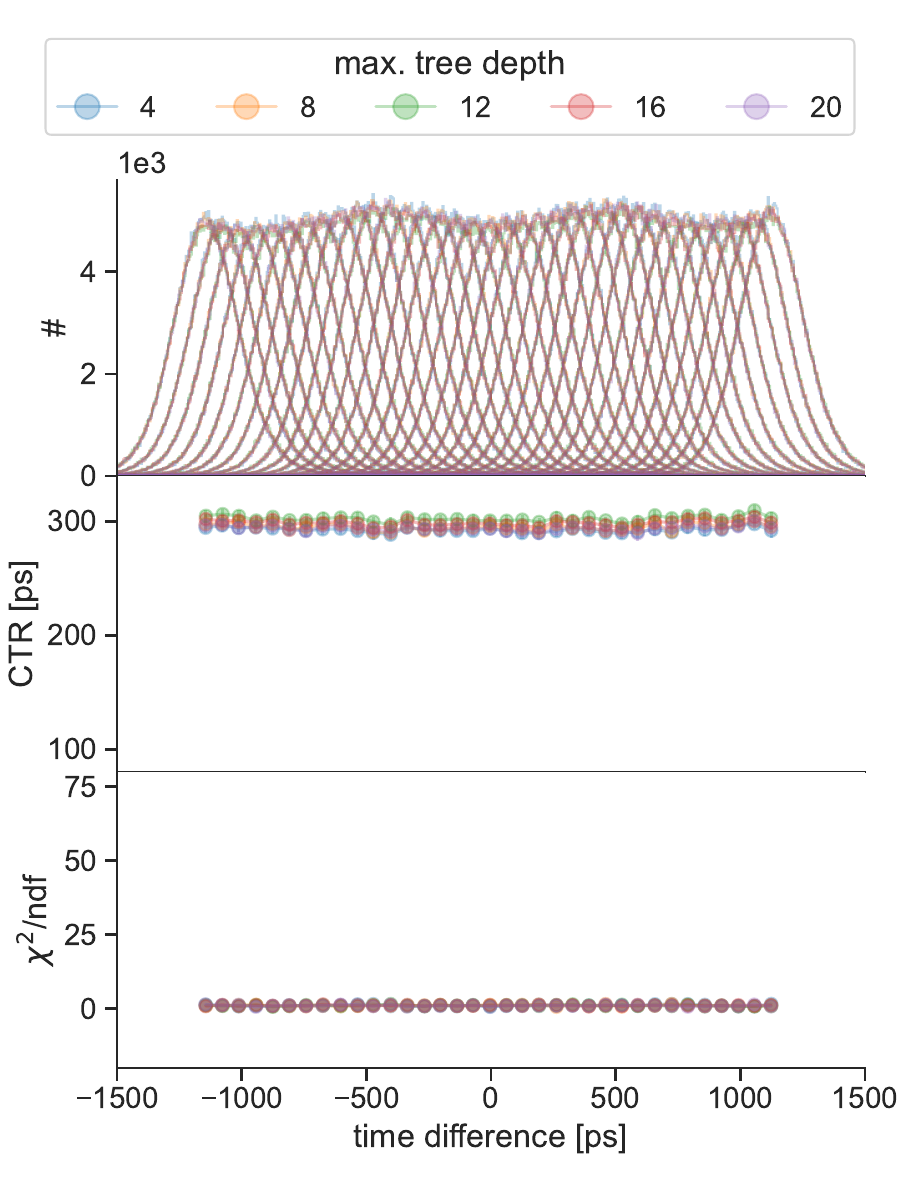}
\caption{\ac{CTR} progression along the $z$-axis of explicit correction models, trained on \qty{100}{\milli \metre} stepping. The upper plot visualizes the indirectly predicted time differences. The middle plot shows the \ac{CTR} value at the specific position. The lower plot visualizes the $\chi^2$/ndf-value received from fitting a Gaussian function to the corresponding time difference distribution. The data shown in this plot was selected to be within an energy window from \qtyrange{430}{590}{\kilo \eV}.}
\label{fig:ctr_z_100}
\end{figure}

All correction models that passed the quality control are able to improve the \ac{CTR}. The achieved timing resolutions estimated from a point source in the isocenter are listed in \cref{tab:ctr430590}, \cref{tab:ctr300700}, and \cref{tab:ctrno}. Furthermore, the results of the two models IM$_{10, 12}$, EM$_{10,4}$, and the results before using machine learning are visualized in \cref{fig:ctr_pred}. The \ac{CTR} progression along the $z$-axis is displayed in \cref{fig:ctr_z}, \cref{fig:ctr_z_100} and \cref{tab:ctr430590z}. All mean \ac{CTR} values are compatible within $1\sigma$ with the estimated \acp{CTR} from the isocenter. While the explicit correction approach remains functional across the full $z$-range, the implicit correction approach is known to experience bias effects at the edges of the training domain (see \cref{fig:example_pred}). This leads to non-Gaussian time distributions, recognizable by the high $\chi^2$/ndf-values that return decreased \acp{CTR} values due to the distribution deformation. In order to provide an unbiased evaluation, the source positions from the gray area are excluded from the evaluation of the implicit correction approach (see \cref{tab:ctr430590z}). It is remarkable that all estimated timing resolutions demonstrate only minor degradation when going to a large energy window, or no energy window at all. The estimated timing resolutions are compatible with each other for a given energy window and correction approach, except for model IM$_{50,8}$. One can observe that the implicit correction approach provides slightly better \ac{CTR} values compared to the explicit correction approach. 

\clearpage

\begin{table}[htb]
\caption{\ac{CTR} values of different training step widths and an energy window of \qtyrange{430}{590}{\kilo \eV}. Models that have not passed the prior quality checks are marked with qcf (quality check failed). The maximal tree depth is denoted with $d$. *This model lies within a transition region and should therefore be interpreted with caution.}
\centering
\begin{tabular}{@{}S[table-format=2.0]ccccccc@{}}
\toprule
 & \multicolumn{7}{c}{CTR [ps]} \\ \midrule
{\multirow{3}{*}{d}} & \multicolumn{2}{c}{\qty{10}{\milli \metre}} & \multicolumn{2}{c}{\qty{50}{\milli \metre}} & \multicolumn{2}{c}{\qty{100}{\milli \metre}} & \multirow{2}{*}{no ML} \\ \cmidrule(lr){2-7}
                   & \cellcolor{\mygray} implicit & explicit & \cellcolor{\mygray} implicit & explicit & \cellcolor{\mygray} implicit & explicit &                      \\ \midrule
4                  & \cellcolor{\mygray} qcf   & $282 \pm 4$   & \cellcolor{\mygray} qcf  & $291 \pm 4$   & \cellcolor{\mygray} qcf		   & $292 \pm 5$   & $371 \pm 6$         \\
8                  & \cellcolor{\mygray} qcf   & $283 \pm 7$   & \cellcolor{\mygray} *$253 \pm 5$   & $292 \pm 3$   & \cellcolor{\mygray} qcf		   & $296 \pm 5$   &                      \\
12                 & \cellcolor{\mygray} $271 \pm 5$   & $291 \pm 4$   & \cellcolor{\mygray} qcf		   & $297 \pm 5$   & \cellcolor{\mygray} qcf		   & $303 \pm 5$   &                      \\
16                 & \cellcolor{\mygray} $280 \pm 4$   & $287 \pm 4$   & \cellcolor{\mygray} qcf		   & $297 \pm 4$   & \cellcolor{\mygray} qcf		   & $300 \pm 4$   &                      \\
20                 & \cellcolor{\mygray} qcf		   & $288 \pm 5$   & \cellcolor{\mygray} qcf		   & $292 \pm 5$   &  \cellcolor{\mygray} qcf           & $293 \pm 5$   &                      \\ \bottomrule
\end{tabular}
\label{tab:ctr430590}
\end{table}

\begin{table}[htb]
\caption{Mean \ac{CTR} values along the $z$-axis of different training step widths and an energy window of \qtyrange{430}{590}{\kilo \eV}. Models that have not passed the prior quality checks are marked with qcf (quality check failed). The maximal tree depth is denoted with $d$. All mean \ac{CTR} values are compatible within $1\sigma$ with the estimated \acp{CTR} from the isocenter (see \cref{tab:ctr430590}).  *This model lies within a transition region and should therefore be interpreted with caution.}
\centering
\begin{tabular}{@{}S[table-format=2.0]ccccccc@{}}
\toprule
 & \multicolumn{7}{c}{CTR [ps]} \\ \midrule
{\multirow{3}{*}{d}} & \multicolumn{2}{c}{\qty{10}{\milli \metre}} & \multicolumn{2}{c}{\qty{50}{\milli \metre}} & \multicolumn{2}{c}{\qty{100}{\milli \metre}} & \multirow{2}{*}{no ML} \\ \cmidrule(lr){2-7}
                   & \cellcolor{\mygray} implicit & explicit & \cellcolor{\mygray} implicit & explicit & \cellcolor{\mygray} implicit & explicit &                      \\ \midrule
4                  & \cellcolor{\mygray} qcf   & $282 \pm 2$   & \cellcolor{\mygray} qcf  & $290 \pm 2$   & \cellcolor{\mygray} qcf		   & $293  \pm 2$   & $371 \pm 2$         \\
8                  & \cellcolor{\mygray} qcf   & $283 \pm 2$   & \cellcolor{\mygray} *$278 \pm 19$   & $290 \pm 2$   & \cellcolor{\mygray} qcf		   & $295  \pm 3$   &                      \\
12                 & \cellcolor{\mygray} $274 \pm 7$   & $290 \pm 2$   & \cellcolor{\mygray} qcf		   & $295 \pm 2$   & \cellcolor{\mygray} qcf		   & $303  \pm 2$   &                      \\
16                 & \cellcolor{\mygray} $275 \pm 6$   & $287 \pm 2$   & \cellcolor{\mygray} qcf		   & $296 \pm 2$   & \cellcolor{\mygray} qcf		   & $299  \pm 2$   &                      \\
20                 & \cellcolor{\mygray} qcf		   & $289 \pm 2$   & \cellcolor{\mygray} qcf		   & $291 \pm 3$   &  \cellcolor{\mygray} qcf           & $294  \pm 2$   &                      \\ \bottomrule
\end{tabular}
\label{tab:ctr430590z}
\end{table}

\clearpage

\subsection{The In-Plane Distribution Study}
\subsubsection{Data Scientific Evaluation}
\begin{figure}[ht]
\centering
\includegraphics[width=0.98\textwidth]{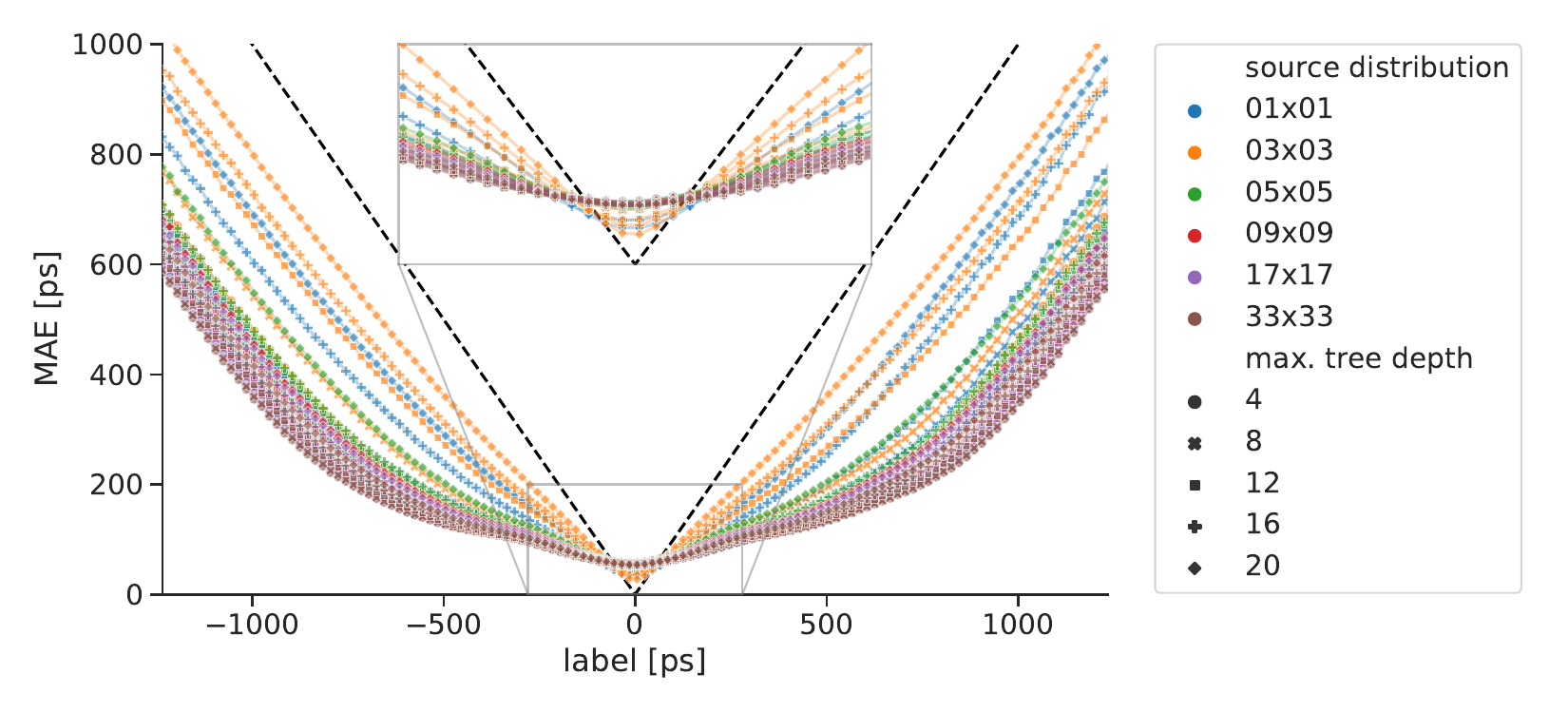}
\caption{\ac{MAE} progression of explicit correction models trained with different maximum tree depth and in-plane source distributions. The dashed black line displays the \ac{MAE} progression of a dummy model predicting for every input a correction value of \qty{0}{\pico \second}.}
\label{fig:mae_study2}
\end{figure}

\begin{figure}[htb]
\centering
\subfloat[Mean \ac{MAE}.]{%
    \includegraphics[width=0.45\textwidth]{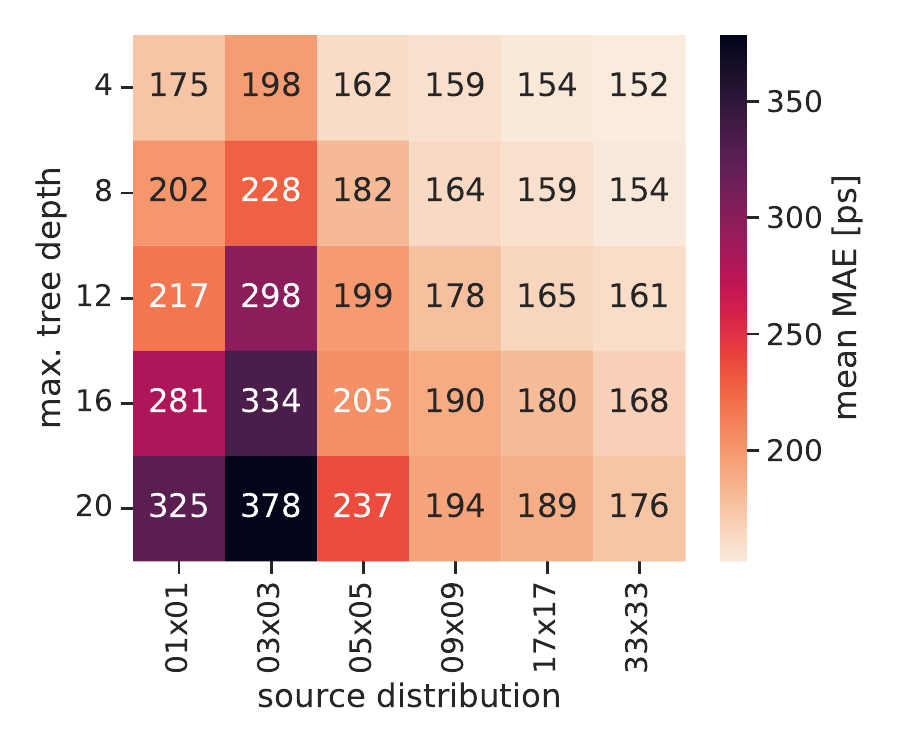}
    \label{fig:mean_mae2}
}
\hfill
\subfloat[Weighted mean \ac{MAE}.]{%
    \includegraphics[width=0.45\textwidth]{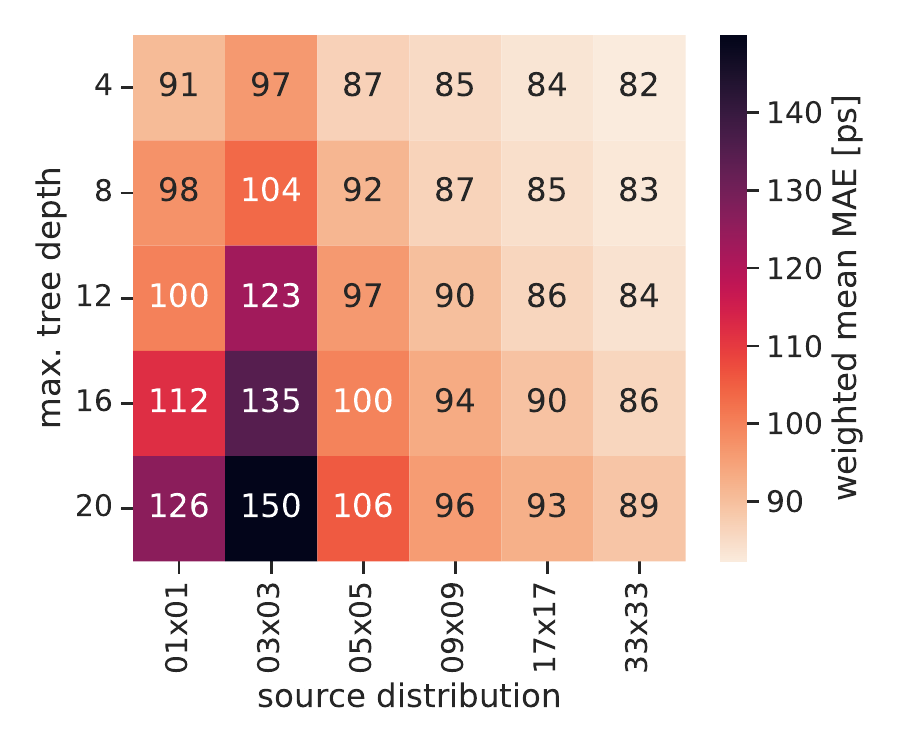}
    \label{fig:weighted_mae2}
}

\caption{Visualizations of the mean \ac{MAE} and weighted mean \ac{MAE} dependent on the maximum tree depth and the source distribution. While the mean \ac{MAE} is given as the mean of the points displayed as a curve in \cref{fig:mae_study2}, the weighted mean is calculated by weighting the \ac{MAE} value of a specific label with the label's occurrence frequency.}
\label{fig:study2_mmae_wmae}
\end{figure}

The data scientific evaluations of the trained explicit correction models are depicted in \cref{fig:mae_study2} and \cref{fig:study2_mmae_wmae}. It can be seen that the \ac{MAE} progression is smooth, and no outliers or oscillation effects are visible. The shape of the \ac{MAE} progression resembles the U-form, which is already known from the previous study. There are slight differences visible for different in-plane source distributions, namely that models trained on extended source distribution show a flatter \ac{MAE} progression than compressed distributions when moving towards label values with a large magnitude. Apparently, this trend occurs for all the different source distributions, except for the case of \numproduct{3x3}, which performs the worst. This trend is reversed for small label magnitudes, where models trained on compressed source distributions show lower \ac{MAE} values. For convenience, \cref{fig:mae_study2} also displays the \ac{MAE} progression of a dummy model predicting a correction value of \qty{0}{\pico \second} for every input.\newline
For better visualization and interpretability, the mean \ac{MAE} values and weighted mean \ac{MAE} values are displayed as heatmaps in \cref{fig:study2_mmae_wmae}. The mean \ac{MAE} values are calculated as the mean of the \ac{MAE} progression curve of \cref{fig:mae_study2}, while the weighted mean \ac{MAE} values are given by considering the relative label occurrence (see \cref{fig:study1_mae10}b)). Both plots reveal the trend of improved \ac{MAE} performance when models with small maximum tree depth are trained on the extended source distribution. Again, the case of \numproduct{3x3} sources serves as an exception.

\subsubsection{Physics-based Evaluation}

\begin{figure}[htb]
\centering
\subfloat[Calculated $s_{\chi^2}$ values.]{%
    \includegraphics[width=0.45\textwidth]{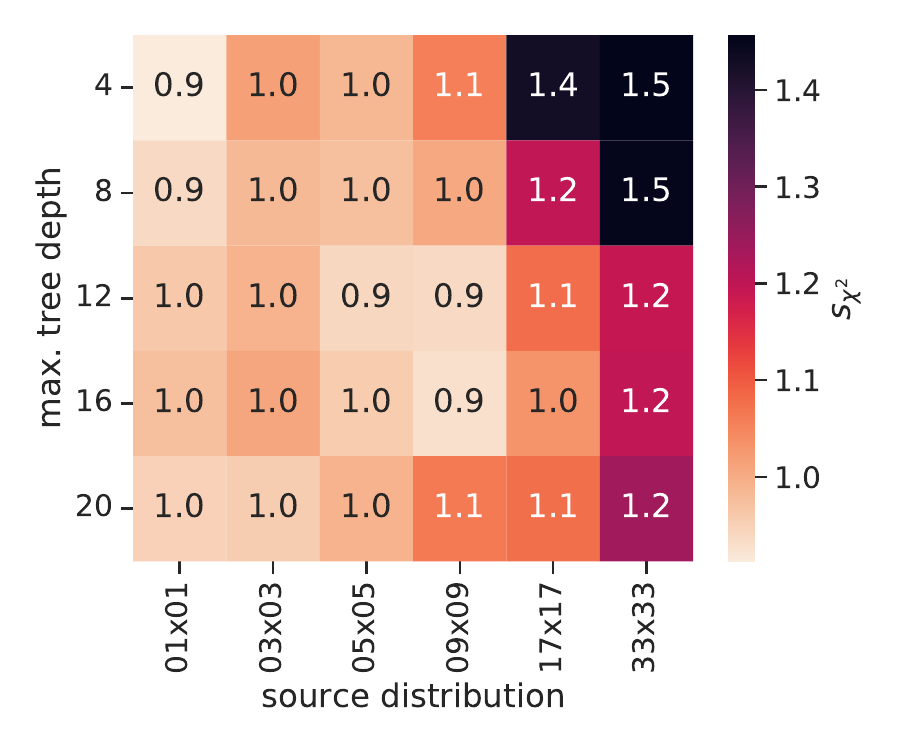}
}
\hfill
\subfloat[Calculated $n_{\overline{\epsilon}}$ values.]{%
    \includegraphics[width=0.45\textwidth]{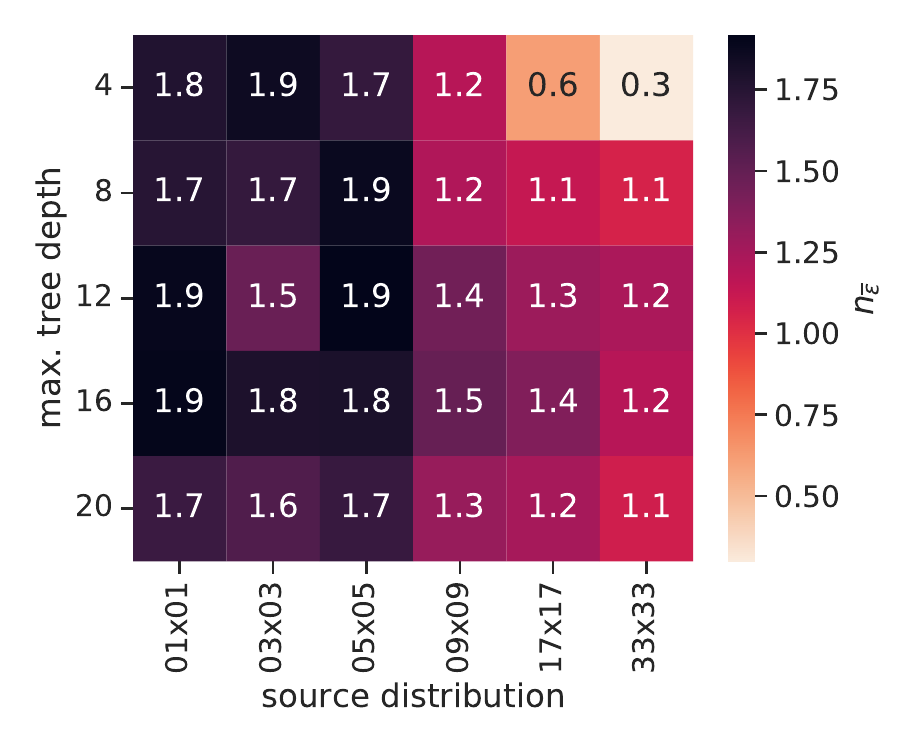}
}

\caption{Visualization of the metrics used during the physics-based evaluation dependent on the maximum tree depth and the source distribution.}
\label{fig:study2_s_and_eps}
\end{figure}

All trained models were evaluated regarding their linearity property and agreement with physics using the methods explained in \cref{subsubsec:physicsEval}. For the sake of better visualization and comparability, the results are displayed as heatmaps showing the $s_{\chi^2}$ and $n_{\overline{\varepsilon}}$ values. In general, the plots reveal that all explicit correction models pass the physics-based evaluation. When looking at the linearity property measure shown in \cref{fig:study2_s_and_eps}a, one can see that models trained on small source distributions tend to have a smaller $\chi^2$-spread than models trained on \numproduct{17x17} sources or more. Regarding different maximum tree depths, no clear trend can be identified.\newline
The heatmap displaying the $\sigma$-environments of the estimated $\overline{\varepsilon}$-value shows an opposing picture. The smallest differences between estimated and theoretic $\overline{\varepsilon}$-value are achieved by models trained on extensive source distributions. Also for this evaluation, no clear tendency regarding the maximum tree depth is visible.\newline
Overall, the plots demonstrate that all explicit correction models show a high agreement with our physics-based expectations with only minor differences between differently trained models.

\subsubsection{PET-based Evaluation}

\begin{figure}[ht]
\centering
\includegraphics[width=\scBroadd \textwidth]{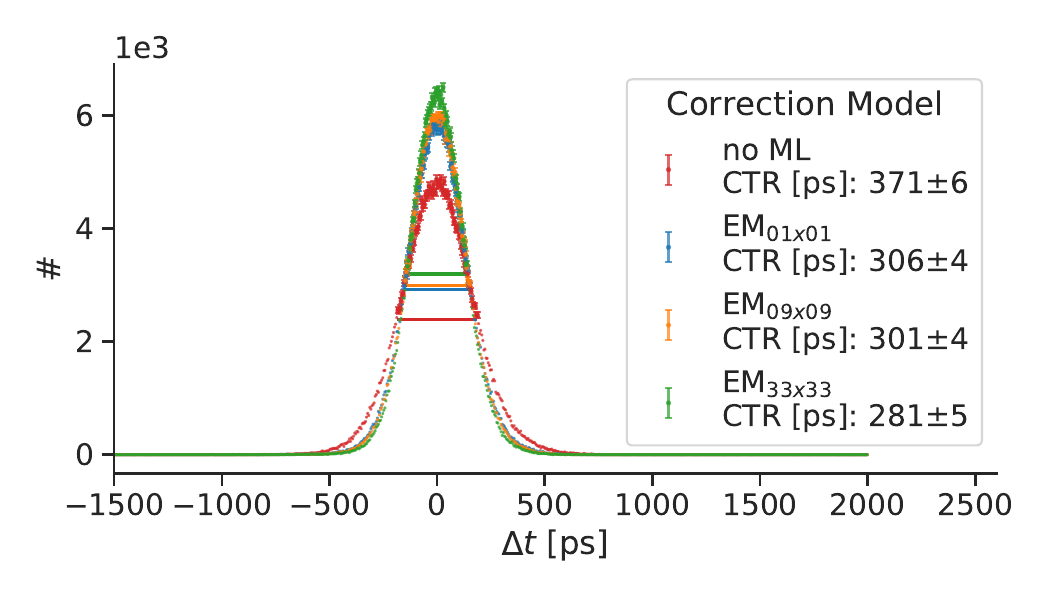}
\caption{Obtained timing resolutions for explicit correction models trained on data with different in-plane source distributions. The values are based on coincidences being in an energy window from \qtyrange{430}{590}{\kilo \eV}. The correction model 'no ML' refers to performing only an analytical time skew calibration (first part of the residual physics calibration scheme), without subsequent use of machine learning.}
\label{fig:ctr_pred_study2}
\end{figure}

The timing resolutions achieved when using the trained explicit correction models are listed in \cref{fig:study2_ctr_430}, \cref{fig:study2_ctr_300}, and \cref{fig:study2_ctr_no}. Some of the resulting time difference distributions are depicted in \cref{fig:ctr_pred_study2}. Models that have been trained on extensive in-plane source distributions show the best \ac{CTR} values. Furthermore, the values indicate the trend that training on more in-plane sources leads to models with a higher ability to correct deteriorating effects, resulting in better \acp{CTR}. An exception is given by models trained on \numproduct{3x3} in-plane sources, which return the worst \ac{CTR} performance. Although no general trend regarding the maximum tree depth is visible, in the two extreme cases (\numproduct{1x1} and \numproduct{33x33}) explicit models with a maximum depth of \num{4} perform the best.\newline
When comparing the \acp{CTR} from coincidences of a small and large energy window, one sees only a minor degradation ($\approx \qty{2}{\percent}$) for a given explicit model compared to the 'no ML' correction ($\approx \qty{20}{\percent}$).\newline
All trained explicit correction models are able to significantly improve the achievable timing resolution compared to performing only an analytical time skew correction without subsequent use of machine learning. For the best case, the \ac{CTR} can be improved by nearly \qty{25}{\percent} for data with an energy from \qtyrange{430}{590}{\kilo \eV}, and \qty{36}{\percent} for data with an energy from \qtyrange{300}{700}{\kilo \eV}.

\begin{table}[htb]
\centering
\caption{Obtained timing resolutions after usage of explicit correction models trained on data with different in-plane source distributions. The values are based on coincidences being in an energy window from \qtyrange{430}{590}{\kilo \eV}. The column 'no ML' refers to performing only an analytical time skew calibration (first part of the residual physics calibration scheme), without subsequent use of machine learning.}
\begin{tabular}{@{}S[table-format=2.0]ccccccc@{}}
\toprule
 & \multicolumn{6}{c}{CTR [ps]} \\ \midrule
{d}  & \cellcolor{\mygray} 01x01       & 03x03       & \cellcolor{\mygray} 05x05       & 09x09       & \cellcolor{\mygray} 17x17       & 33x33  & \cellcolor{\mygray} no ML     \\ \midrule
4  & \cellcolor{\mygray} $306 \pm 4$ & $328 \pm 5$ & \cellcolor{\mygray} $300 \pm 5$ & $301 \pm 4$ & \cellcolor{\mygray} $289 \pm 4$ & $281 \pm 5$ & \cellcolor{\mygray} $371 \pm 6$ \\
8  & \cellcolor{\mygray} $316 \pm 4$ & $334 \pm 5$ & \cellcolor{\mygray} $312 \pm 5$ & $300 \pm 5$ & \cellcolor{\mygray} $294 \pm 4$ & $282 \pm 7$ & \cellcolor{\mygray} \\
12 & \cellcolor{\mygray} $305 \pm 5$ & $341 \pm 5$ & \cellcolor{\mygray} $310 \pm 5$ & $299 \pm 4$ & \cellcolor{\mygray} $290 \pm 5$ & $288 \pm 4$ & \cellcolor{\mygray} \\
16 & \cellcolor{\mygray} $313 \pm 5$ & $340 \pm 5$ & \cellcolor{\mygray} $303 \pm 5$ & $299 \pm 4$ & \cellcolor{\mygray} $293 \pm 6$ & $287 \pm 5$ & \cellcolor{\mygray} \\
20 & \cellcolor{\mygray} $322 \pm 5$ & $345 \pm 4$ & \cellcolor{\mygray} $306 \pm 5$ & $294 \pm 3$ & \cellcolor{\mygray} $291 \pm 5$ & $287 \pm 5$ & \cellcolor{\mygray} \\ \bottomrule
\end{tabular}
\label{fig:study2_ctr_430}
\end{table}

\begin{table}[htb]
\centering
\caption{Obtained timing resolutions after usage of explicit correction models trained on data with different in-plane source distributions. The values are based on coincidences being in an energy window from \qtyrange{300}{700}{\kilo \eV}. The column 'no ML' refers to performing only an analytical time skew calibration (first part of the residual physics calibration scheme), without subsequent use of machine learning.}
\begin{tabular}{@{}S[table-format=2.0]ccccccc@{}}
\toprule
 & \multicolumn{6}{c}{CTR [ps]} \\ \midrule
{d}  & \cellcolor{\mygray} 01x01       & 03x03       & \cellcolor{\mygray} 05x05       & 09x09       & \cellcolor{\mygray} 17x17       & 33x33  & \cellcolor{\mygray} no ML     \\ \midrule
4  & \cellcolor{\mygray} $314 \pm 4$ & $337 \pm 4$ & \cellcolor{\mygray} $309 \pm 3$ & $308 \pm 3$ & \cellcolor{\mygray} $295 \pm 3$ & $287 \pm 4$ &\cellcolor{\mygray} $446 \pm 5$ \\
8  & \cellcolor{\mygray} $323 \pm 4$ & $345 \pm 4$ & \cellcolor{\mygray} $323 \pm 4$ & $307 \pm 4$ & \cellcolor{\mygray} $301 \pm 3$ & $289 \pm 5$ &\cellcolor{\mygray} \\
12 & \cellcolor{\mygray} $311 \pm 5$ & $360 \pm 3$ & \cellcolor{\mygray} $319 \pm 4$ & $308 \pm 3$ & \cellcolor{\mygray} $300 \pm 3$ & $295 \pm 3$ &\cellcolor{\mygray} \\
16 & \cellcolor{\mygray} $330 \pm 3$ & $368 \pm 4$ & \cellcolor{\mygray} $313 \pm 4$ & $308 \pm 4$ & \cellcolor{\mygray} $300 \pm 4$ & $295 \pm 3$ &\cellcolor{\mygray} \\
20 & \cellcolor{\mygray} $349 \pm 5$ & $383 \pm 4$ & \cellcolor{\mygray} $320 \pm 2$ & $303 \pm 3$ & \cellcolor{\mygray} $299 \pm 3$ & $294 \pm 4$ &\cellcolor{\mygray} \\ \bottomrule
\end{tabular}
\label{fig:study2_ctr_300}
\end{table}

\section{Discussion}
In this work a novel formulation of a residual physics-based timing calibration was introduced, capable of providing explicit timestamp corrections values. Two distinguished aspects of the calibration were investigated. In the transaxial performance study the effect of reduced source positions along the $z$-axis in comparison to the established implicit correction approach was evaluated. The in-plane distribution study analyzed the effect of different source distributions located in a plane on the explicit correction models .\newline
The data scientific evaluation of the transaxial performance study revealed both approaches are capable of significantly improving the achievable \ac{CTR} when training and testing data have the same spatial sampling. All explicit and implicit correction models showed a smooth \ac{MAE} progression, except for the implicit correction model with a maximum tree depth of \num{20}. Although the learning curves of this model showed no abnormal course and converged to a minimum, the validation error was clearly higher than for the other implicit models. We assume that the model might be too complex for the given problem, suppressing effective learning. As soon as the spatial sampling of the training data became more sparse than the spatial sampling of the test data, implicit correction models showed the tendency to oscillation effects. Those effects led to a very good performance at positions that are known to the model, and worse performance at unknown positions. It can be stated that implicit models with a high maximum tree depth tended to be more sensitive to undersampling than models with less complexity. Although the \ac{MAE} curves of the explicit correction models seemed steeper, indicating a higher \ac{MAE}, one has to consider the underlying label distribution. While there are many training samples with low label magnitudes, the number of high magnitude label samples is low making it hard for the models to learn characteristic patterns. The results suggested that explicit correction models are more robust against a spatially undersampled training dataset. This can be reasoned by the fact, that the label distribution is independent from the location and number of source positions along the $z$-axis. The physics-based evaluation confirmed the previously mentioned findings, namely that implicit correction models show a dependence on the spatial sampling of the training data. Furthermore, the results demonstrate that the dependence on the source location is removed for explicit correction models. This implies that explicit approach does not demand a dedicated motorized calibration setup, which significantly increases the practicability minding an in-system application. The robustness of the explicit correction models is reasoned by the way the residuals are defined. The corresponding mathematical considerations can be found in the appendix. Furthermore, we want to underline that the linearity evaluation for the explicit correction models was performed on the full test data range. Since the implicit correction models show large bias effects at the edges of the test range, a meaningful linearity analysis could only be performed on roughly \qty{60}{\percent} of the test data range. Minding spatially undersampled training data, no strong dependence on the maximum tree depth was found for explicit correction models contrary to implicit correction models. The \ac{PET}-specific evaluation revealed that models from both correction approaches which passed the quality checks were able to significantly improve the timing performance of the used \ac{PET} detectors. The estimated \ac{CTR} values from the isocenter are in good agreement with the mean \ac{CTR} archived on the allowed $z$-range. The best results were achieved by the implicit correction models. The explicit correction models show a slightly worse performance in \ac{CTR}, which might be neglectable considering the scale of the improvements. Both correction approaches demonstrate only small \ac{CTR} degradation when enlarging the energy window, which suggests that both methods are able to correct timewalk effects successfully. This suggestion is supported by an extensive feature importance analysis conducted in a prior study \cite{naunheim_improving_2023} using implicit correction models.\newline
In the in-plane distribution study, we demonstrated that the explicit correction formulation is capable of proving good results, even if the training data does not have multiple source positions along the $z$-axis. Furthermore, we investigated which effects occur if the in-plane source distribution is compressed to one point source or extended to a quasi-continuous distribution. Regarding those two extreme cases, we had an application for a full \ac{PET} scanner in mind, where one could potentially use a point source in the isocenter or a thin phantom filled with activity. During the data scientific evaluation, all models showed a similar \ac{MAE} progression, although models trained on extensive in-plane source distributions showed slightly better performance, especially at large label magnitudes. Since one also has to consider that the overall training statistics do strongly differ (e.g., the number of training samples for the \numproduct{1x1} case is approximately a factor $33^2$ smaller compared to the \numproduct{33x33} case), we cannot strictly identify if this effect comes from the source distribution or statistics. An indication that the performance depends on the in-plane arrangement of the sources might be the case where \numproduct{3x3} sources were utilized since those models performed worse compared to the \numproduct{1x1} case even though the training data was bigger by a factor of \num{9}. However, in the \numproduct{3x3} case, it must also be said that the sources are oriented more towards the detector edge than towards the detector center, which can be a further influencing factor. The physics-based evaluation showed that the linearity property is stronger pronounced in models trained on comprised source distribution, while the $\varepsilon$-agreement is slightly higher for models trained on extended source distributions. Overall, all explicit correction models passed the physics-based quality check. The \ac{CTR} values being achieved by the trained explicit correction models are significantly better (nearly \qty{25}{\percent} for data with an energy from \qtyrange{430}{590}{\kilo \eV}, and \qty{36}{\percent} for data with an energy from \qtyrange{300}{700}{\kilo \eV}) compared to not using machine learning. Furthermore, only a small performance degradation in \ac{CTR} is observed for a large energy window, which suggests that trained models are capable of correcting time walk effects, which is interesting when high sensitivity is to be considered. All trained models show a consistent picture of the achievable timing resolution for a given in-plane source distribution. Minding the two extreme cases of having only one source or $33^2$ sources, the results suggest that models with a low maximum tree depth are preferable for the explicit formulation. Our previous studies \cite{naunheim_improving_2023} showed that higher maximum tree depth resulted in the best performance for the implicit correction approach. These two findings are not mutually exclusive or contradictory. Instead, we believe that the implicit correction approach is better suited for boosting models, where iteratively weak learners are added to minimize the residuals of the existing models. Minding the label distribution in the explicit correction case, from a statistical standpoint a good first correction estimation would be \qty{0}{\pico \second}. For the implicit correction approach, a first estimation being \qty{0}{\pico \second} would be unsuitable for many samples, thus the model has to be more complex. Since the explicit correction approach yields promising results even with small maximum tree depth, it becomes possible to reduced the model size by a factor of $2^{18-4}$ (see \cref{eq:MR} and minding the best model of \cite{naunheim_improving_2023}). This massive reduction in memory makes the approach interesting for an on-chip application \cite{krueger_high-throughput_2023} and edge-AI application were a high data throughput is required.

\section{Summary \& Outlook}

In this work, we presented a novel way of defining timing residuals using a residual physics-based timing calibration approach, allowing explicit access to \ac{TOF} corrections. We demonstrated that the explicit correction approach offers many benefits compared to the implicit correction models: independence from the used spatial sampling along the axis in transaxial direction, high linearity across the full test data range, and only minor degradation regarding the achievable \ac{CTR}. Since the novel formulation does not rely on measuring source positions between facing detectors, we removed the demand for a dedicated motorized setup, making the method more practical for an in-system application. Compared to our proof-of-concept study, where the best implicit correction model had a maximum tree depth of \num{18} \cite{naunheim_improving_2023}, the novel explicit approach offers a significant reduction in the memory requirements of a model by a factor of approximately $2^{18-4}$ (see \cref{eq:MR}) which makes it suitable for high throughput applications like a \ac{PET} scanner. \newline
For the future, and with the perspective of an in-system application, we want to test how stable the correction models perform when applied to different detector stacks of the same design and material. Through the design of our features, we expect some degree of robustness. However, we are also investigating foundational modeling approaches \cite{masbaum_first_2024,lavronenko_towards_2024} that will hopefully allow us to train generalistic models suitable for many detector stacks of the same kind. Additionally, we plan to investigate how various in-plane source distributions affect the training of correction models, with the goal of designing a suitable calibration phantom. Furthermore, we want to explore in future studies how the explicit correction models perform concerning the reported feature importance when applied repeatedly to data corrected with the predicted corrections.

\section{Acknowledgment}
Open access funding provided by the Open Access Publishing Fund of RWTH Aachen University.\newline
We thank the RWTH Aachen University Hospital workshop, Harald Radermacher, and Oswaldo Nicolas Martinez Arriaga for the help when setting up the translation stage system.

\newpage
\section*{Appendix}
\section*{Consideration of the Linearity of Explicit Correction Models}

Although, we will not provide a formal proof why the explicit correction models show a strong linearity robustness, in this section we analyze the behavior more theoretically. We will follow the notation used in the previous sections.\newline
Let $\{t_{a,i}\}$ and $\{t_{b,i}\}$ be the set of measured timestamps of detector $a$ and $b$. By using the explicit correction approach, our model generates predictions $\{p_i\}$ serving as correction values for the measured timestamps such that the corrected timestamps $\{t_{a,i}^{\text{corr}}\}$, $\{t_{b,i}^{\text{corr}}\}$ are given as
\begin{equation}
\begin{aligned}
t_{a,i}^{\text{corr}} &= t_{a,i} - p_i, \\
t_{b,i}^{\text{corr}} &= t_{b,i} + p_i,
\end{aligned}
\end{equation}
and the corrected time difference $\Delta t_i^{\text{corr}}$ is given as
\begin{equation}
\Delta t_i^{\text{corr}} = t_{a,i}^{\text{corr}} - t_{b,i}^{\text{corr}}.
\end{equation}
The linearity analysis relies on a linear regression using

\begin{equation}
\Delta t_{\mathbb{E}} \approx \mu (z; \varepsilon, b) = - \frac{2}{c} \cdot \varepsilon \cdot z + b,
\label{eq:linReg_app}
\end{equation}
with the variables on the right side of the equation sign being defined in \cref{eq:linReg}. For our consideration we want to take a closer look on left side of the equation sign, since it is the part affected by the explicit correction model. Formally the expected time difference is given as
\begin{equation}
\Delta t_{\mathbb{E}} = \mathbb{E} \left[ \{ \Delta t_i \} \right],
\end{equation}
which translates in the case of explicitly corrected timestamps to

\begin{align}
\Delta t_{\mathbb{E}} &= \mathbb{E} \left[ \{ \Delta t_i^{\text{corr}} \} \right] \\
 &= \mathbb{E} \left[ \{ t_{a,i}^{\text{corr}} - t_{b,i}^{\text{corr}} \} \right] \\
 &= \mathbb{E} \left[ \{ \left( t_{a,i} - p_i \right) - \left( t_{b,i} + p_i \right) \} \right] \\
 &= \mathbb{E} \left[ \{ \left( t_{a,i} - t_{b,i} \right) - 2 p_i  \} \right] \\
 &= \mathbb{E} \left[ \{ \Delta t_i - 2 p_i  \} \right].
 \label{eq:app_insert}
\end{align}

If we assume that the explicit correction model was successfully trained, we can approximate the predictions by the labels (see \cref{eq:explicitLabel}),

\begin{equation}
p_i \approx l_i \coloneq \frac{\Delta t_{m,i} - \Delta t_{\mathbb{E}} (z)}{2}.
\label{eq:explicitLabel_study2}
\end{equation}

Inserting \cref{eq:explicitLabel_study2} into \cref{eq:app_insert}, yields
\begin{align}
\Delta t_{\mathbb{E}} &= \mathbb{E} \left[ \biggl\{ \Delta t_i - 2 \cdot \left( \frac{\Delta t_i - \mathbb{E}\left[ \{ \Delta t_i \} \right] }{2} \right) \biggr\} \right] \\
&= \mathbb{E} \Bigl[ \Bigl\{ \mathbb{E} \left[ \{ \Delta t_i \} \right] \Bigr\} \Bigr] \\
&= \mathbb{E} \left[ \{ \Delta t_i \} \right],
\end{align}
which is the expected time difference for the non-corrected timestamps. Although, we assumed that the predictions can be approximated by the labels, the presented considerations provide some basic understanding of the robustness of the explicit correction models. Furthermore, the requirements and assumptions can probably a bit softened since we approximate the expectation value by a Gaussian fit.

\newpage

\begin{table}[htb]
\caption{\ac{CTR} values of different training step widths and an energy window of \qtyrange{300}{700}{\kilo \eV}. Models that have not passed the prior quality checks are marked with qcf (quality check failed). The maximal tree depth is denoted with $d$. *This model lies within a transition region and should therefore be interpreted with caution.}
\centering
\begin{tabular}{@{}S[table-format=2.0]ccccccc@{}}
\toprule
 & \multicolumn{7}{c}{CTR [ps]} \\ \midrule
{\multirow{3}{*}{d}} & \multicolumn{2}{c}{\qty{10}{\milli \metre}} & \multicolumn{2}{c}{\qty{50}{\milli \metre}} & \multicolumn{2}{c}{\qty{100}{\milli \metre}} & \multirow{2}{*}{no ML} \\ \cmidrule(lr){2-7}
                   & \cellcolor{\mygray} implicit & explicit & \cellcolor{\mygray} implicit & explicit & \cellcolor{\mygray} implicit & explicit &                      \\ \midrule
4                  & \cellcolor{\mygray} qcf   & $289 \pm 3$   & \cellcolor{\mygray} qcf   & $297 \pm 4$   & \cellcolor{\mygray} qcf           & $302 \pm 3$   & $446 \pm 5$         \\
8                  & \cellcolor{\mygray} qcf   & $291 \pm 4$   & \cellcolor{\mygray} *$277 \pm 3$  & $300 \pm 4$   & \cellcolor{\mygray} qcf           & $305 \pm 4$   &                      \\
12                 & \cellcolor{\mygray} $284 \pm 4$   & $298 \pm 5$   & \cellcolor{\mygray} qcf		   & $306 \pm 3$   & \cellcolor{\mygray} qcf           & $311 \pm 4$   &                      \\
16                 & \cellcolor{\mygray} $288 \pm 4$   & $292 \pm 4$   & \cellcolor{\mygray} qcf		   & $305 \pm 2$   & \cellcolor{\mygray} qcf           & $310 \pm 3$   &                      \\
20                 & \cellcolor{\mygray} qcf		   & $296 \pm 4$   & \cellcolor{\mygray} qcf		   & $300 \pm 2$   & \cellcolor{\mygray} qcf           & $299 \pm 6$   &                      \\ \bottomrule
\end{tabular}
\label{tab:ctr300700}
\end{table}

\begin{table}[]
\caption{\ac{CTR} values of different training step width and no energy filter. Models that have not passed the prior quality checks are marked with qcf (quality check failed). The maximal tree depth is denoted with $d$. *This model lies within a transition region and should therefore be interpreted with caution.}
\centering
\begin{tabular}{@{}S[table-format=2.0]ccccccc@{}}
\toprule
 & \multicolumn{7}{c}{CTR [ps]} \\ \midrule
{\multirow{3}{*}{d}} & \multicolumn{2}{c}{\qty{10}{\milli \metre}} & \multicolumn{2}{c}{\qty{50}{\milli \metre}} & \multicolumn{2}{c}{\qty{100}{\milli \metre}} & \multirow{2}{*}{no ML} \\ \cmidrule(lr){2-7}
                   & \cellcolor{\mygray} implicit & explicit & \cellcolor{\mygray} implicit & explicit & \cellcolor{\mygray} implicit & explicit &                      \\ \midrule
4                  & \cellcolor{\mygray} qcf   & $317 \pm 3$   & \cellcolor{\mygray} qcf   & $327 \pm 3$   & \cellcolor{\mygray} qcf           & $332 \pm 3$   & $674 \pm 5$         \\
8                  & \cellcolor{\mygray} qcf   & $320 \pm 3$   & \cellcolor{\mygray} *$312 \pm 2$   & $330 \pm 3$   & \cellcolor{\mygray} qcf           & $335 \pm 4$   &                      \\
12                 & \cellcolor{\mygray} $314 \pm 3$   & $327 \pm 3$   & \cellcolor{\mygray} qcf		   & $336 \pm 2$   & \cellcolor{\mygray} qcf           & $347 \pm 3$   &                      \\
16                 & \cellcolor{\mygray} $317 \pm 3$   & $321 \pm 3$   & \cellcolor{\mygray} qcf		   & $338 \pm 3$   & \cellcolor{\mygray} qcf           & $346 \pm 3$   &                      \\
20                 & \cellcolor{\mygray} qcf		   & $326 \pm 3$   & \cellcolor{\mygray} qcf		   & $331 \pm 2$   & \cellcolor{\mygray} qcf           & $334 \pm 4$   &                      \\ \bottomrule
\end{tabular}
\label{tab:ctrno}
\end{table}

\begin{table}[]
\centering
\caption{Obtained timing resolutions after usage of explicit correction models trained on data with different in-plane source distributions. No restrictions on the energy are applied. The column 'no ML' refers to performing only an analytical time skew calibration (first part of the residual physics calibration scheme), without subsequent use of machine learning.}
\begin{tabular}{@{}S[table-format=2.0]ccccccc@{}}
\toprule
 & \multicolumn{6}{c}{CTR [ps]} \\ \midrule
{d}  & \cellcolor{\mygray} 01x01       & 03x03       & \cellcolor{\mygray} 05x05       & 09x09       & \cellcolor{\mygray} 17x17       & 33x33  & \cellcolor{\mygray} no ML     \\ \midrule
4  & \cellcolor{\mygray} $349 \pm 3$ & $392 \pm 4$ & \cellcolor{\mygray} $338 \pm 3$ & $337 \pm 2$ & \cellcolor{\mygray} $324 \pm 3$ & $316 \pm 3$ &\cellcolor{\mygray} $674 \pm 5$ \\
8  & \cellcolor{\mygray} $361 \pm 2$ & $403 \pm 4$ & \cellcolor{\mygray} $358 \pm 3$ & $336 \pm 4$ & \cellcolor{\mygray} $331 \pm 3$ & $318 \pm 4$ &\cellcolor{\mygray} \\
12 & \cellcolor{\mygray} $344 \pm 3$ & $446 \pm 4$ & \cellcolor{\mygray} $358 \pm 3$ & $340 \pm 3$ & \cellcolor{\mygray} $328 \pm 4$ & $323 \pm 3$ &\cellcolor{\mygray} \\
16 & \cellcolor{\mygray} $392 \pm 3$ & $473 \pm 4$ & \cellcolor{\mygray} $350 \pm 3$ & $342 \pm 3$ & \cellcolor{\mygray} $334 \pm 3$ & $325 \pm 3$ &\cellcolor{\mygray} \\
20 & \cellcolor{\mygray} $444 \pm 4$ & $521 \pm 4$ & \cellcolor{\mygray} $369 \pm 3$ & $336 \pm 3$ & \cellcolor{\mygray} $332 \pm 3$ & $324 \pm 3$ &\cellcolor{\mygray} \\ \bottomrule
\end{tabular}
\label{fig:study2_ctr_no}
\end{table}

\begin{table}[htb]
\centering
\caption{Settings used during measurement.}
\begin{tabular}{@{}l S@{}}
\toprule
Parameter                & {Value}                        \\ \midrule
measurement mode             & {hardware coincidence trigger} \\
coincidence window (hardware) & 3 \\
coincidence window (software) [\unit{\nano \second}]      & 50                           \\
breakdown voltage [\unit{\volt}]       & 32                           \\
overvoltage        [\unit{\volt}]      & 7                            \\
vth\_t1, vth\_t2, vth\_e & {20, 20, 15}                   \\ \bottomrule
\end{tabular}
\label{tab:measurement_settings}
\end{table}

\clearpage
\bibsetup
\printbibliography

\end{document}